\documentclass[aps,prl,reprint,onecolumn,superscriptaddress,nofootinbib]{revtex4-2}

\usepackage[left=1.2in,right=1.2in,top=1in,bottom=1in]{geometry}
\usepackage{graphicx}
\usepackage{xcolor}
\usepackage{lmodern}
\usepackage{amsmath,amssymb,amsfonts}
\usepackage{braket}
\usepackage{svg}
\usepackage{subfig}
\usepackage{adjustbox}
\usepackage{comment}
\usepackage[colorlinks=true,allcolors=red!60!black]{hyperref}

\begin{document}

\title{Localized Decoherence as a Constructive Tool: An Optical Simulation of Matter-Wave Interference}

\author{Hadis Torbatiyan}
\thanks{These two authors contributed equally}
\affiliation{ University of Ljubljana, Faculty of Mathematics and Physics, Jadranska 19, Ljubljana, Slovenia}

\author{Bedir Halcı}
\thanks{These two authors contributed equally}
\affiliation{ University of Ljubljana, Faculty of Mathematics and Physics, Jadranska 19, Ljubljana, Slovenia}

\author{Adrian Udovi\v{c}ić}
\affiliation{ University of Ljubljana, Faculty of Mathematics and Physics, Jadranska 19, Ljubljana, Slovenia}

\author{Lara Ul\v{c}akar} 
\affiliation{ University of Ljubljana, Faculty of Mathematics and Physics, Jadranska 19, Ljubljana, Slovenia}
\affiliation{Jo\v{z}ef Stefan Institute, Jamova 39, Ljubljana, Slovenia}

\author{Rainer Kaltenbaek}
\email{rainer.kaltenbaek@fmf.uni-lj.si}
\affiliation{ University of Ljubljana, Faculty of Mathematics and Physics, Jadranska 19, Ljubljana, Slovenia}

\begin{abstract}
Decoherence is conventionally regarded as a detrimental process that suppresses quantum coherence and limits the performance of quantum technologies. 
Contrary to this 
view, we show that spatially localized decoherence can instead be exploited as an interferometric tool. Beyond its conceptual implications, localized decoherence provides a novel approach to prepare arbitrary macroscopic spatial superpositions for testing quantum physics with massive particles. 
We demonstrate the proof of principle by making use of the correspondence between optics and matter-waves through the equivalence of the respective propagation functions. By employing stochastic phase fields to the optical counterpart to model the decoherence, our experiments confirm the expected interference. 
The results establish localized decoherence as a resource for quantum state engineering.
\end{abstract}

\maketitle

\section{Introduction}
The decoherence of quantum systems arising from their coupling to the environment~\cite{Joos1985a,Zurek2003a} generally suppresses the coherence required to observe quantum phenomena.
As quantum technologies continue to progress towards increasingly large and complex systems, understanding and controlling decoherence is quickly becoming a technological, as well as a fundamental challenge. 
Conventional descriptions of decoherence often assume that the coupling between the system and its environment is spatially homogeneous~\cite{Schlosshauer2019a}. 
For extended quantum systems, however, environmental interactions generally vary across the wavefunction, causing different spatial domains to decohere differently. 
Understanding the consequences of such localized decoherence is therefore becoming increasingly important for the development of large-scale quantum technologies.

Matter-wave interferometry provides a powerful platform to address this challenge by probing quantum coherence in massive particles and its degradation through interactions with the environment. 
Early experiments with fullerene molecules characterized decoherence arising from collisions with background gas and from thermal radiation~\cite{Hornberger2003b,Hackermueller2004a}. 
Subsequent advances in interferometer design and particle preparation have enabled interference with increasingly massive particles. In particular, Talbot-type interferometers using optical phase gratings~\cite{Gerlich2007, Hornberger2009a} or optical absorptive gratings~\cite{Nimmrichter2011a,Haslinger2013a,Eibenberger2013a,Fein2019a,Pedalino2026a} have allowed matter-wave interference to be observed while avoiding interactions between material gratings~\cite{Hornberger2003b,Hackermueller2004a,Brezger2002a} and the test particles. 
These techniques have achieved matter-wave interferometry with high-mass particles up to $\sim 10^2\ \mathrm{kDa}$~\cite{Pedalino2026a}. 

As the test mass increases, the required optical wavelength decreases. Eventually, this leads to a single scattering event being able to localize the test particle and destroy the quantum superposition~\cite{Nimmrichter2013b,Belenchia2019a}.
A fundamentally different perspective suggested to harness localized decoherence to prepare macroscopic superpositions, using decoherence itself as an interferometric tool~\cite{Kaltenbaek2012b}. 
There, it was suggested to apply a tightly focused laser pulse to a portion of an expanding matter wave, locally inducing decoherence while leaving the remainder of the wave packet coherent. During subsequent free evolution, the coherent domains would then overlap and interfere. 

In the present paper, we demonstrate this idea by placing it in a theoretical framework and confirming it experimentally through an optical simulation. Our approach exploits a correspondence defined through the propagation description of matter waves and classical optical beams. For the experiments performed, we consider two different geometries: a tightly focused beam creating a single radially symmetric decoherence domain, and two overlapping beams creating a periodic decoherence profile. In the optical counterpart, the decoherence is modelled as dephasing, which is realized with a spatial light modulator (SLM). 
The results of our work are twofold: we demonstrate that localized decoherence can generate interference phenomena, and we show that this can be harnessed as an interferometric tool for preparing arbitrary spatial superpositions of massive particles.

\section{Localized strong decoherence due to light scattering}
\label{sec:MWI}

We consider particles, where the scattering of light is the dominant cause of decoherence.
Decoherence can be understood as a loss of phase relations~\cite{Zurek2003a}. Therefore, it can be modelled by introducing random phase shifts to constituents of the wavefront.
Here, we consider the case of strong decoherence as suggested in Ref.~\cite{Kaltenbaek2012b}, where the phase relations are completely lost.
We introduce a
heuristic model~\footnote{see Eq.~4.37 in Ref.~\cite{Zurek2003a}.}, where the localized decoherence is represented by a stochastic phase field applied on $\Omega$, which denotes a domain of spatial points where scattering occurred and the particle got decohered probabilistically. Each individual point in $\Omega$  obtains a random phase shift $\vartheta_{xy}$ that belongs to a random variable $\Theta_{xy}$, uniformly distributed over the interval $[0,2\pi)$. The resulting stochastic wavefunction can be described  as~\cite{Stern1990, Dalibard1992} 
\begin{equation}
\Psi(x,y) = \sqrt{p}\ e^{i\vartheta_{xy}}\ket{\mathrm{\Omega}} + \sqrt{1-p}\ \ket{\mathrm{\Omega}^\complement},
\label{eq:psi}
\end{equation}
where the state $\ket{\Omega}$ represents the states in $\Omega$, and $\ket{{\Omega}^\complement}$ in the complementary domain $\Omega^\complement$. The stochastic wavefunction $\Psi$ exists in the Hilbert space $\mathcal{H} = \mathcal{H}^\Omega \oplus \mathcal{H}^{\Omega^\complement}$ consisting of two respective sub-domains of identical cardinality. These properties justify the use of superpositions of states in separate sub-domains.
The normalization parameter $p$ depends on 
how the sub-domains and corresponding spatial regions are related to each other through our description of decoherence.

The physical density matrix is obtained 
after ensemble averaging over random phase realizations. Since $\mathbf{E}[e^{i\vartheta_{xy}}]=0$, the result is of the form:
\begin{equation}
\rho = \mathbf{E}[\Psi\Psi^*] = p\,\rho_{\mathrm{dec}} + (1-p)\,\rho_{\mathrm{coh}},
\label{eq:density}
\end{equation}
where $\rho_{\mathrm{dec}}$ corresponds to the decohered part of the matter-wave, whereas $\rho_{\mathrm{coh}}$ describes the unperturbed part,
reproducing the density matrix expected in the strong-decoherence limit.

In the following, we present 
proposals for methods that harness local decoherence due to scattering of light in the context of matter-wave experiments. All proposals are based on a nanoparticle, initially trapped and cooled to its motional ground state. The particle is released from the trap, so its wavefunction can expand for some time. The expansion time is selected such that the extent of the wavefunction will be significantly larger than the wavelength of the scattered light.

\subsection{Single-Domain Local Decoherence}
\label{ssec:sdld}
In this method, the decohering pulse is radially symmetric. After the interaction with the decohering pulse, the quantum system can be described by the stochastic wavefunction in Eq.~\ref{eq:psi}. The particle evolves again freely for some additional time. During this evolution, the coherent portions of the system will continue expanding and eventually overlap. As a result, interference fringes can emerge in the final spatial distribution.

\subsection{Periodic Domain Local Decoherence}
\label{ssec:pdld}
The concept of localized decoherence using laser scattering can be expanded beyond radially symmetric geometries. We propose to create periodically modulated decoherence patterns by illuminating the particle with two coherent laser beams with a beam waist $w_\mathrm{dec}$, a wavelength $\lambda_\mathrm{dec}$ and intersecting at an angle $\varphi$.
The resulting local scattering probability will follow an  accordion-like intensity distribution~\cite{Cruickshank2026}. We expect the subsequent matter-wave evolution to create a Talbot carpet~\cite{Case2009}.

\section{Correspondence \& strong decoherence models}
\label{sec:corresp}
A direct experimental realization of localized decoherence as proposed in Sec.~\ref{sec:MWI} requires advanced experimental infrastructure, stringent isolation from environmental decoherence, and precise control of massive quantum systems.
To overcome these challenges and to provide an experimentally accessible platform for investigating localized decoherence, we define a correspondence between the free evolution of a matter wave of mass $m$ in time, and the paraxial propagation of a classical optical beam with wavelength $\lambda$ in space. Both are solutions to Helmholtz equations in their respective coordinates, propagating according to their respective Green's function formulation~\cite{Brukner1997,Santos2018}\footnote{compare Eq. (2) in \cite{Brukner1997} and Eq. (4) in \cite{Santos2018}.}. The equivalence between the Green's functions for a matter wave and a classical optical beam through a coordinate transform results in a combined compact correspondence as:
\begin{equation}\label{eq:plcorr}
    \frac{m}{\hbar}\frac{\rho^2}{2t}= \frac{2\pi}{\lambda}\frac{r^2}{2z},
\end{equation}
where $\{\rho,t\}$ and $\{r,z\}$ are the cylindrical coordinates of the propagating matter wave and optical beam, respectively. This results in a scaling between the two coordinate systems. An 
unscaled version of the correspondence was mentioned before \cite{Deng1999}.

A full description of the evolution of matter waves in the presence of localized decoherence still requires mapping the decoherence into an equivalent phase modulation of the electromagnetic field. This can be achieved in the form of dephasing, formulated according to Eq.~\ref{eq:psi}, by introducing an SLM.
Using this dephasing-based modelling of decoherence, every pixel of the SLM is assigned either to the decohered domain ($\Omega$) or to the complementary coherent domain ($\Omega^\complement$).
The pixels belonging to $\Omega$ acquire spatially uncorrelated random phases, causing decoherence in the corresponding portion of the beam.
The pixels in the complementary domain are set to a constant phase, which keeps that portion of the beam coherent.
All the discussed geometries give rise to interference patterns. The correspondence remains valid during subsequent propagation after the decoherence because matter waves and electromagnetic fields are both governed by the same propagator mechanics~\cite{Feynman1948,Santos2018}.

For the decoherence geometries defined in Sec.~\ref{ssec:sdld} and Sec.~\ref{ssec:pdld}, we consider three representative models:
\begin{itemize}
\item Opaque Disk (OD). A circular region of radius $a$ is defined in the centre of the SLM. The pixels inside the region, corresponding to $\Omega$, are assigned independent random phases uniformly distributed over the interval $[0, 2\pi)$, while all pixels outside the region, $\Omega^\complement$, are assigned zero phase. This configuration represents the simplest model of localized decoherence, where full decoherence is guaranteed within an effective radius. The domain splitting corresponds to a spatial division.

We consider this model to represent the matter wave getting decohered by a beam containing a high number of photons, where each single photon is able to cause full decoherence. The theoretical expectation for the created intensity pattern, derived in Appendix~\ref{app:od},
is of the form:
\begin{equation}\label{eq:odf}
   I_\mathrm{OD}(r,z) \propto \left|e^{-i\Gamma_r/2}+i\frac{k}{z}\int_0^a\rho e^{i\Gamma_\rho/2}J_0(kr\rho /z)d\rho\right|^2,
\end{equation}
where $\Gamma_j=kj^2/z$.
This gives rise to an Arago spot in the  centre $I_\mathrm{OD}(0,z) \propto 1$.

\item Gaussian Opacity (GO). The probability that a given pixel is phase randomized follows a radially symmetric 2D Gaussian envelope with standard deviation $\sigma$. For every pixel, we generate a random number uniformly distributed on the interval $[0,1]$ and compare it with a corresponding Gaussian threshold. If the random number lies below that threshold, corresponding to $\Omega$, the pixel is assigned a phase chosen uniformly from the interval $[0,2\pi)$; the contrary outcome corresponds to $\Omega^\complement$, and the phase is set to zero. Therefore, the domain splitting does not correspond to a spatial division.

This represents the case where a lower number photons are present in the beam that is decohering the matter wave and every single photon is able to cause full decoherence, but the occurrence of decoherence is probabilistic. The intensity behaviour, derived in Appendix~\ref{app:go}, is of the form:
\begin{equation}\label{eq:gof}
    I_\mathrm{GO}(r,z) \propto 1
    +\frac{\Gamma_\sigma^2}{1+\Gamma_\sigma^2}e^{-\Gamma_r\frac{\Gamma_\sigma}{1+\Gamma_\sigma^2}}
    -2\frac{\Gamma_\sigma^2}{1+\Gamma_\sigma^2}e^{-\frac{\Gamma_r}{2}\frac{\Gamma_\sigma}{1+\Gamma_\sigma^2}}\mathrm{c\Gamma_\sigma^{-1}s}\left(\frac{\Gamma_r}{2}\frac{1}{1+\Gamma_\sigma^2}\right),
\end{equation}
where $\mathrm{c\Gamma_\sigma^{-1}s}(x)=\cos(x) + \Gamma_\sigma^{-1}\sin(x)$.
This produces a shadowing effect at the centre $I_\mathrm{GO}(0,z) \propto 1/(1+k^2\sigma^4/z^2)$.

    \item Periodic Opacity (PO). Similarly to GO, the occurrence of decoherence is probabilistic at a given spatial point, but the threshold definition is done through an accordion-like function instead, which depends on the three parameters $w$, $l$, and $\varphi$. A Talbot carpet is expected to be observed, with the Talbot length $z_\mathrm{T}=2\varsigma^2/\lambda$, where the grid separation is $\varsigma=l/2\sin(\varphi/2)$. The effective number of stripes in the grid, denoted by $n$, can be determined from the number of accordion peaks that have a modulation probability greater than a threshold probability that we chose to be $1/2$, resulting in $n =2(w/l)\sqrt{2\ln(2)}\tan(\varphi/2)$. The corresponding derivation and analysis are provided in Appendix~\ref{app:po}.
    
    Here, the parameter $\varphi$ is equal to the angle between the lasers described in \ref{ssec:pdld}. The parameters $w$ and $l$, on the other hand, need to be transformed from their matter-wave counterparts $w_\mathrm{dec}$ and $\lambda_\mathrm{dec}$. The relation is provided via Eq.~\eqref{eq:plcorr} by choosing a Talbot time $t_\mathrm{T}$, which is the characteristic time required for the formation of the Talbot carpet in the matter wave counterpart,  resulting in $w/w_\mathrm{dec}=l/\lambda_\mathrm{dec}=\sqrt{(z_\mathrm{T}/t_\mathrm{T})(m\lambda/h)}$.
\end{itemize}

\section{Experimental realizations via optical simulation}
\label{sec:optical}

As a proof of principle, we optically analyse a simplified version of the correspondence discussed in Sec.~\ref{sec:corresp}. To this end, we investigate a zoomed-in patch of the wavefront, corresponding to a plane wave. 
The observation of an interference pattern for this simple case justifies the expectation of interference also for a Gaussian wavefront.

\subsection{Implementation}

\begin{figure}[h]
\centering  

\includegraphics[width=0.8\textwidth]{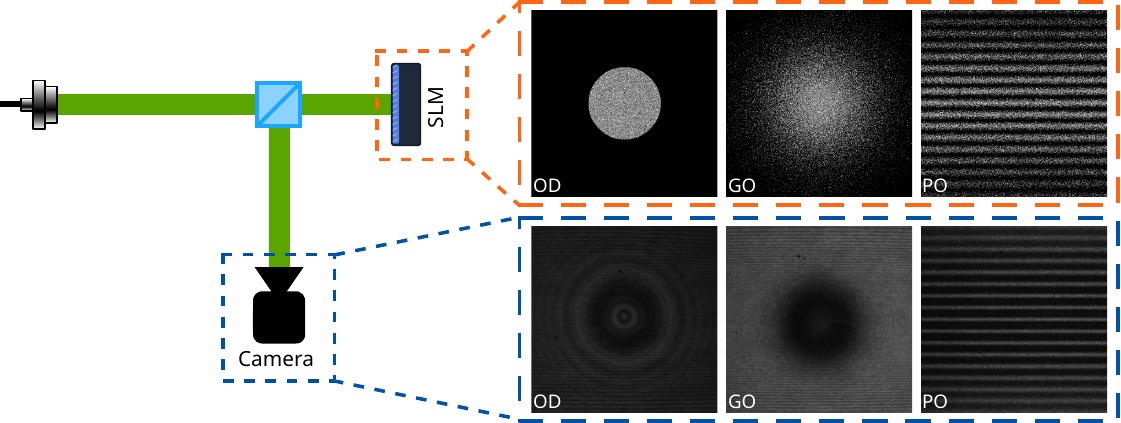}
  \caption{Sketch of the experimental setup for optical simulations of local decoherence. The spatial light modulator (SLM) encodes random phase fields on the incident beam. The phase fields are depicted inside the red dashed area and follow three different random distributions: Opaque Disk (OD), Gaussian Opacity (GO) and Periodic Opacity (PO). 
  Corresponding examples of camera frames are shown inside the blue dashed area.  
  The low-visibility horizontal stripes apparent in the OD and GO cases result from diffraction due to the rectangular aperture of the SLM.
  The cropped SLM patterns and camera frames both cover areas of $4.12\ \mathrm{mm}\times 4.12\ \mathrm{mm}$.}
\label{fig:setup}
\end{figure}

Fig.~\ref{fig:setup} illustrates the optical setup used to simulate spatially dependent decoherence. 
For our experiment, we used a continuous-wave laser (Toptica DL pro) operating at a wavelength of $\lambda = 519\,\mathrm{nm}$ and an output power of {$P_0=\,65.3\,\mu\mathrm{W}$}. 
The optical beam was collimated and we chose a radius of $w_0 \approx 1\,\mathrm{cm}$ to make it larger than the active area of the SLM. The incident wavefront can then be approximated as a plane wave.
We use a non-polarizing beam splitter (BS) that transmits the beam onto the SLM at direct incidence.  After reflection from the SLM, the beam propagates back through the BS. 
The reflected half of the beam was analysed using a Neptune-C II camera mounted on a translation rail to allow millimetre-precision movements of the camera along the beam. The camera had a resolution of $2712\times1538$ pixels with a pixel size of $2.9\ \mathrm{\mu m}\times2.9\ \mathrm{\mu m}$. 

To implement phase modulation, we used a Holoeye PLUTO 2.1 phase-only SLM with an active area of $1920\times1080$ pixels and a pixel size of $8.0\ \mathrm{\mu m}\times8.0\ \mathrm{\mu m}$. 
The maximum refresh rate of the device was $60\,\mathrm{Hz}$, which also determined the rate at which the random phase patterns were updated. 
The incident polarization was adjusted to ensure phase-only modulation. 
We  confirmed the absence of residual amplitude modulation using a Michelson interferometer.
Using this setup, we also identified a range of pixel values over which the phase response of the SLM was linear.
The phase distributions displayed on the SLM were encoded as 8-bit greyscale images, where the greyscale range $[40,230]$ maps to phase shifts spanning the interval $[0,2\pi)$.
This corresponds to a phase resolution of $32.2\ \mathrm{mrad}$ per unit greyscale.
Examples of the phase patterns used to simulate different cases of local decoherence are shown in Fig.~\ref{fig:setup}. In the case of OD and GO, there exist horizontal stripes with low-visibility. This is due to the finite-size of the SLM, creating rectangular diffraction. We suppressed this effect significantly by introducing a smoothening mask of randomised pixels at the edges, which is shown in Appendix~\ref{app:mask}.

\subsection{Single-domain local decoherence}
\label{ssec:sdldexp}

Here, we present the results for single-domain local decoherence, which simulates the interaction between the expanded wave front of a massive particle and a Gaussian beam.
The dephasing encoded on the SLM is described by the OD and GO models, as illustrated in Figure~\ref{fig:setup}.
Examples of frames captured by the camera are provided in the same figure.

\begin{figure}[h]
 \centering  
\includegraphics[width=1\textwidth]{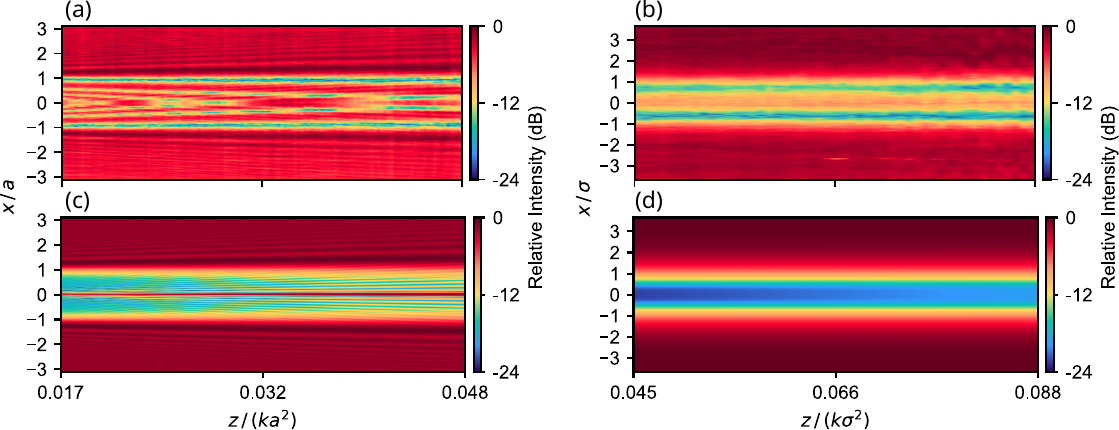}
 \caption{Panels (a) and (c) show the experimental data and the corresponding theoretical expectations, respectively, for the interference patterns generated with the spatial light modulator (SLM) for encoding the Opaque Disk (OD) model  with a size parameter $a=\,800\ \mathrm{\mu m}$.
 Panels (b) and (d) show the experimental data and the theoretical expectations, respectively, for the Gaussian Opacity (GO) model with $\sigma=\,679\ \mathrm{\mu m}$.
 The intensity pattern in the horizontal plane containing the beam axis is shown in logarithmic colour scale as a function of the distance from the SLM. The logarithmic scale is chosen to allow for easier comparability. Each slice at a given $z$ is a single-pixel horizontal slice at the centre of the camera images.}
\label{fig:Arago}
\end{figure}

The intensity patterns in the  horizontal plane containing the beam axis are shown in Fig.~\ref{fig:Arago}.
We compare experimental results for the case of OD and GO shown in (a) and (b), respectively, with the corresponding theoretical predictions shown in (c) and (d), which were obtained from equations \eqref{eq:odf} and \eqref{eq:gof}.
The modulation sizes for the two cases were $a=\,800\ \mathrm{\mu m}$ and $\sigma=\,679\ \mathrm{\mu m}$.
In the case of OD in Fig.~\ref{fig:Arago}(a), we observe oscillations between constructive and destructive interference inside the geometric shadow, while interference fringes are present outside the geometric shadow.
In contrast, the theoretical prediction shown in Fig.~\ref{fig:Arago}(c), features an unperturbed Arago spot behaviour. The intensity pattern generated by GO is shown in Fig.~\ref{fig:Arago}(b). In agreement with theory, there are no interference fringes at this scale;\footnote{see Fig.~\ref{fig:gosim} for the full scale to see the interference fringes.} however, while  theory predicts a complete shadowing in the decohered domain, the experimental results show an increase in the central intensity.

There exist several reasons for these discrepancies. Apart from the finite size of the SLM that causes rectangular diffraction, the main contribution here is that the SLM is not a perfect phase modulator. 
A portion of the light is reflected without modulation and stays coherent with the rest of the light.
This $\pi$-shifted light leads to possible modulations in the intensity pattern. Any deviation from the $\pi$ phase shift can cause additional disturbances. All these contributions are discussed and shown in detail in Appendix~\ref{app:fits} with experimental data.

\subsection{Periodic-domain local decoherence}\label{ssec:pdldexp}

We present the results for periodic decoherence, where the  dephasing encoded on the SLM is described by the PO model. The decoherence distribution along the $x$-axis is given by an accordion intensity profile. To observe the Talbot carpet in the experimentally available range, we selected $z_\mathrm{T}=25$ cm. To ensure that the Talbot carpet is not destroyed over a distance $2z_\mathrm{T}$ by the expansion due to the finite number of stripes, we selected $n=8$ stripes. Choosing a Talbot time of $t_\mathrm{T} = 120\ \mathrm{ms}$ for the matter wave counterpart, realised with a test mass of $m=500\ \mathrm{kDa}$, decohered using laser beams with beam waist $w_\mathrm{dec} = 2\ \mathrm{\mu m}$ and wavelength $\lambda_\mathrm{dec} = 200\ \mathrm{nm}$, the corresponding PO parameters are $w=2.33\ \mathrm{mm}$, $l=232.7\ \mathrm{\mu m}$ and $\varphi=54.36^{\circ}$.

\begin{figure}[h]
 \centering  

\includegraphics[width=0.9\textwidth]{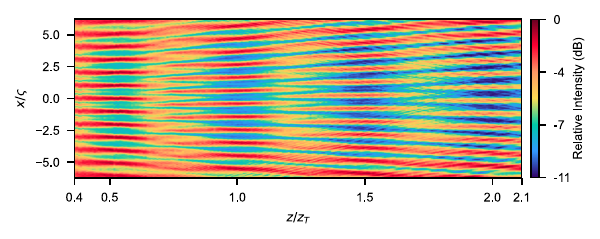}
  \caption{Interference pattern generated with the spatial light modulator (SLM) encoding  the Periodic Opacity (PO) model with the grid separation $\varsigma = 254.4\,\mu\mathrm{m}$ and Talbot length $z_\mathrm{T}=25\,$cm. The intensity pattern in the vertical plane containing the beam axis  is shown in logarithmic colour scale as a function of distance from the SLM. Each slice at a given $z$ is a single-pixel vertical slice at the centre of the camera images.}
\label{fig:Talbot}
\end{figure}

The implementation of the decoherence distribution profile of the SLM is illustrated in Figure~\ref{fig:setup}. An example of a frame captured by the camera is provided in the same figure. The intensity pattern in the vertical plane containing the beam axis, expressed in units of Talbot length, is shown in Fig.~\ref{fig:Talbot}. At $z=z_\mathrm{T}$ and $z=2z_\mathrm{T}$, the observed pattern shows a revival of the periodicity of the grid created on the SLM~\footnote{up to the expansion due to the grid being finite}. At $z=z_\mathrm{T}/2$ and $z=3z_\mathrm{T}/2$, the maxima and minima of the interference pattern are reversed.

\section{Discussion}
\label{sec:disc}

Here, we interpret the experimental results given in Sec.~\ref{sec:optical}. 
First, we discuss the presence and evolution of interference fringes in the case of radially symmetric models in more detail, then we discuss the implications of the experimental results for the corresponding matter-wave case.

The interference fringes observed in the intensity patterns shown in Figs.~\ref{fig:Arago}(c) and \ref{fig:Arago}(d) are heavily dependent on the spatial distribution of the studied decoherence. To investigate this dependence, we studied a generalised Gaussian opacity model (GGO), where the smoothness of the transition between the two domains can be adjusted parametrically. The results show that the hardness of the boundary affects the position and visibility of the interference fringes. 
We investigated this experimentally with a non-physical model, denoted as circular boundary (CB), where a $\pi$-phase shifted disk region is separated from the surrounding region. Our observations confirm the dominant influence of the hard boundary, given that the created interference fringes outside of the geometric shadow are exactly the same as those observed in OD case.
The GGO and CB models, and the experimental data related to them, are discussed in detail in Appendix~\ref{app:addmod}.

The optical simulations can be directly interpreted in terms of the corresponding matter-wave dynamics through the established correspondence. In particular, using Eq.~\eqref{eq:plcorr}, the relation between the optical propagation distance  $\Delta z$ and the corresponding evolution time of a matter-wave $\Delta t$ can be written as $\Delta t = ({m\lambda}/{h})({\delta^2}/{\mu^2}) \Delta z $.
Here,  $\delta$ is the radial size of the decohered domain in the matter-wave experiment, and $\mu$ is the corresponding modulation size in the optical simulation.

For the optical experiments that we conducted with the PO model (Sec. \ref{ssec:pdldexp}), the corresponding Talbot time is $120\ \mathrm{ms}$ for the selected test mass of $500\ \mathrm{kDa}$.  Because the corresponding time scale increases linearly with the particle mass, the observation of the full Talbot evolution becomes increasingly demanding for more massive particles. In the case of OD, the Arago spot half-way develops at a distance equal to the disk radius $a$.
That is why we already see the presence of the Arago spot in the centre for the optically tested distances $z\gg a$, with a radius $a=0.8\,\mathrm{mm}$ (Sec. \ref{ssec:sdldexp}). A corresponding matter-wave experiment can be implemented with the same evolution time and decoherence size $\delta$ as in the Talbot case, but with a particle mass of $m=18.5\ \mathrm{MDa}$. This is nearly two orders of magnitude greater than the mass achievable in the Talbot case, allowing matter-wave experiments with significantly more massive particles.

\section{Conclusions}
\label{sec:concl}
The optical simulation experiments conducted and the studied strong decoherence models show that an interference pattern can be generated via local decoherence.
Our proof-of-principle demonstration implemented this decoherence via the local dephasing of a planar optical wavefront. 
The subsequent optical propagation corresponds directly to the evolution of a zoomed-in patch of a matter wave. 
 
For the radially symmetric models of decoherence we investigated, we observed the appearance of interference fringes, depending on the smoothness of the boundary between the decohered and the coherent domains. In the case of an accordion-like periodic modulation of decoherence, we observed the formation of a Talbot carpet. Our results provide a possible route towards using decoherence as a tool for preparing and manipulating matter waves. More broadly, we showed that the spatial distribution of decoherence can serve not only as a limitation but also as a resource for quantum-state engineering and matter-wave interferometry.

\section*{Acknowledgments}
We thank Igor Muševič and Miha Škarabot for lending us the spatial light modulator. RK thanks Nikolai Kiesel for initial discussions about localized decoherence and about optical simulations of matter-wave interference, and we thank Žiga Pušavec for helpful discussions about the experimental setup. We thank Erik Gajić for his work in the initial preparations of the experimental setup. The authors acknowledge the use of ChatGPT (OpenAI) to assist with improving the accessibility of the manuscript, with debugging and modifying Python code used to generate patterns on the spatial light modulator and for data visualization. The authors take full responsibility for the scientific content, analysis, interpretation, and conclusions of the manuscript.

\section*{Data Availability}
Data and code are made available at Zenodo~\cite{Torbatiyan2026b},
and additional data are available from the corresponding author upon request.

\section*{Funding information}
RK acknowledges support by the Slovenian Research and Innovation Agency (ARIS) under contracts no. P1-0416, N1-0501.
All authors acknowledge financial support from ARIS through the UL VIP project (KTTK21) under contract no. SN-ZRD/22-27/510.
LU acknowledges funding by the Slovenian Research Agency under contract no. P1-0044. AU, HT, and BH acknowledge funding by ARIS under contract no. P1-0416. LU, RK, BH, AU were supported by the Republic of Slovenia (MVZI) and the European Union - NextGenerationEU (SiQUID-101091560).

 \section*{Author Contributions}
RK devised the original idea. HT and AU worked on the initial preparation of the experimental setup. All authors contributed to choosing the set of experiments to be performed. BH calculated the set of experimental parameters. RK suggested the initial version of the GO model. HT, AU and LU performed initial experiments. BH suggested the final version of the GO model. LU suggested the CB model. BH suggested the OD and GGO models. BH developed the theory for the scaled correspondence, the stochastic wavefunction description, and the modeling of strong decoherence as point-wise probabilistic dephasing. BH suggested the use of smoothening mask. HT performed the optical simulation experiments presented in the manuscript. HT processed the raw data. HT and BH analysed the data. HT prepared the figures present in the manuscript and the Appendices. All authors contributed to the discussion and interpretation of the data.  LU wrote the first draft of the manuscript. All authors contributed to writing the manuscript equally. BH wrote the first draft of the Appendices. BH (lead), LU and RK wrote the final version of the Appendices. LU and RK supervised the project.

\appendix

\section{Modelling of strong decoherence}\label{app:models}

The mathematical descriptions of the two radially symmetric models described in the paper is given below. All the models are governed by the Fresnel integral, since the experimental point of operation is in the near field. The general formulation of the electric field resulting from the subsequent propagation after modulation of a plane wave $E_0$ is of the form $E_j(r,\theta)=E_0e^{ikz}e^{i\Gamma_r/2}u_j(r,\theta)$, where we introduced $\Gamma_j=kj^2/z$ as a shorthand notation, and the corresponding spatial amplitude $u_j(r,\theta)$ is:
\begin{equation}\label{eq:fresnel}
    u_j\left(r,\theta \right)=\frac{k}{2\pi iz}\int _0^{2\pi }\int _0^{\infty}M_j\left(\rho ,\phi \right)e^{i\Gamma_\rho/2}e^{-i\frac{kr\rho}{z}\cos \left(\phi-\theta \right)}\rho d\rho d\phi, 
\end{equation}
where $(r,\theta)$ is the coordinate system at the imaging plane, $(\rho,\phi)$ is the coordinate system at the modulation plane, $z$ is the linear distance between the two planes, and $M_j(\rho,\phi)$ describes point-wise modulation for decoherence distribution model $j$. The corresponding intensity pattern is given by the expected value operator, similar to the case of a matter wave described by Eq.~\eqref{eq:psi} for the stochastic wavefunction, $I_j(r,\theta) \propto \mathbf{E}\left[|u_j(r,\theta)|^2\right]$.

We define the subsequent formulations for the domain splitting of decoherence by using a probability field and corresponding probability thresholds. Our description is similar to the probabilistic formulation of stochastic trajectories~\cite{Carmichael1993}\footnote{see Eq. (7.31) given under Sec. 7.5, Lecture 7 in Ref.~\cite{Carmichael1993}}.

\subsection{Opaque disk (OD) model}\label{app:od}

A basic model for the distribution of the decoherence is to define an effective radius of the decohering laser beam, which we denote as the beam radius, and apply it as a hard phenomenological boundary, within which the phase information is fully lost.
One can mathematically formulate this using a partial function in the form:
\begin{equation}\label{eq:od_basic_tr}
    M_{\mathrm{OD}}\left(\rho ,\phi \right) = \left\{0\le \rho <a: e^{i\alpha _{\rho \phi }},a\le \rho:1\right\},
\end{equation}
where $\alpha _{\rho \phi }$ are instances of a random variable $A_{\rho \phi }\sim \mathcal{U}_{\left[-\pi ,\pi \right)}$. This method is illustrated as a procedure in column (a) of Fig. \ref{fig:ODGO}.
Note that the beam radius $a$ is effectively a hard phenomenological boundary between the dephased and the unperturbed domains, meaning the domain splitting corresponds to a spatial division.

We denote this as the ``opaque disk'' model to reflect how the random phase region effectively corresponds to an opaque region. By evolving the beam using the diffraction integral given in Eq.~\eqref{eq:fresnel}, for the point-wise modulation function given in Eq.~\eqref{eq:od_basic_tr}, the first part of the radial integral in the interval $[0, a)$ vanishes when evaluating the azimuthal integral. The surviving part yields:
\begin{equation}
    u_{\mathrm{OD}}\left(r,\theta \right)=\frac{k}{iz}
    \int _a^{\infty}\rho e^{i\Gamma_\rho/2}J_0\left(\frac{kr\rho}{z}\right) d\rho,
\end{equation}
resulting in the intensity pattern
\begin{equation}\label{eq:od_the}
    I_\mathrm{OD}(r,z) \propto \left|e^{-i\Gamma_r/2}+i\frac{k}{z}\int_0^a\rho e^{i\Gamma_\rho/2}J_0\left(\frac{kr\rho}{z}\right)d\rho\right|^2,
\end{equation} 
giving rise to the well-known Arago-spot behaviour $I_\mathrm{OD}(0,z) \propto 1$ for the case of a plane wave incident field.

\begin{figure}[h]
 \centering  
 \includegraphics[width=0.95\textwidth]{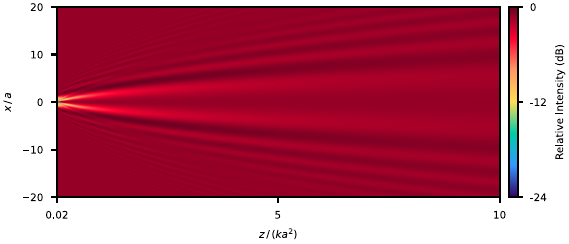}
 \caption{Theoretically expected interference pattern for the Opaque Disk (OD) model given in Eq.~\eqref{eq:od_the}.  The intensity pattern in the horizontal plane containing the beam axis is shown in logarithmic colour scale as a function of the distance from the spatial light modulator (SLM). The logarithmic scale is chosen for comparability.
 }
\label{fig:odsim}
\end{figure}

The expected full scale intensity pattern for the OD model, containing both the near and the far field cases, is shown in Fig.~\ref{fig:odsim}. The interference fringes are present both in the near and far field.

\subsection{Gaussian opacity (GO) model}\label{app:go}

The OD model lacks the effect of a probabilistic rescaling of the decoherence in the modulation region. It guarantees that any photon that is within the region will always decohere the wave-packet locally. However, considering the matter-wave counterpart, a more realistic description is that the probability for a decoherence event to occur should decrease with an increasing distance from the centre.

\begin{figure}[h]
 \centering  
 \includegraphics[width=0.9\textwidth]{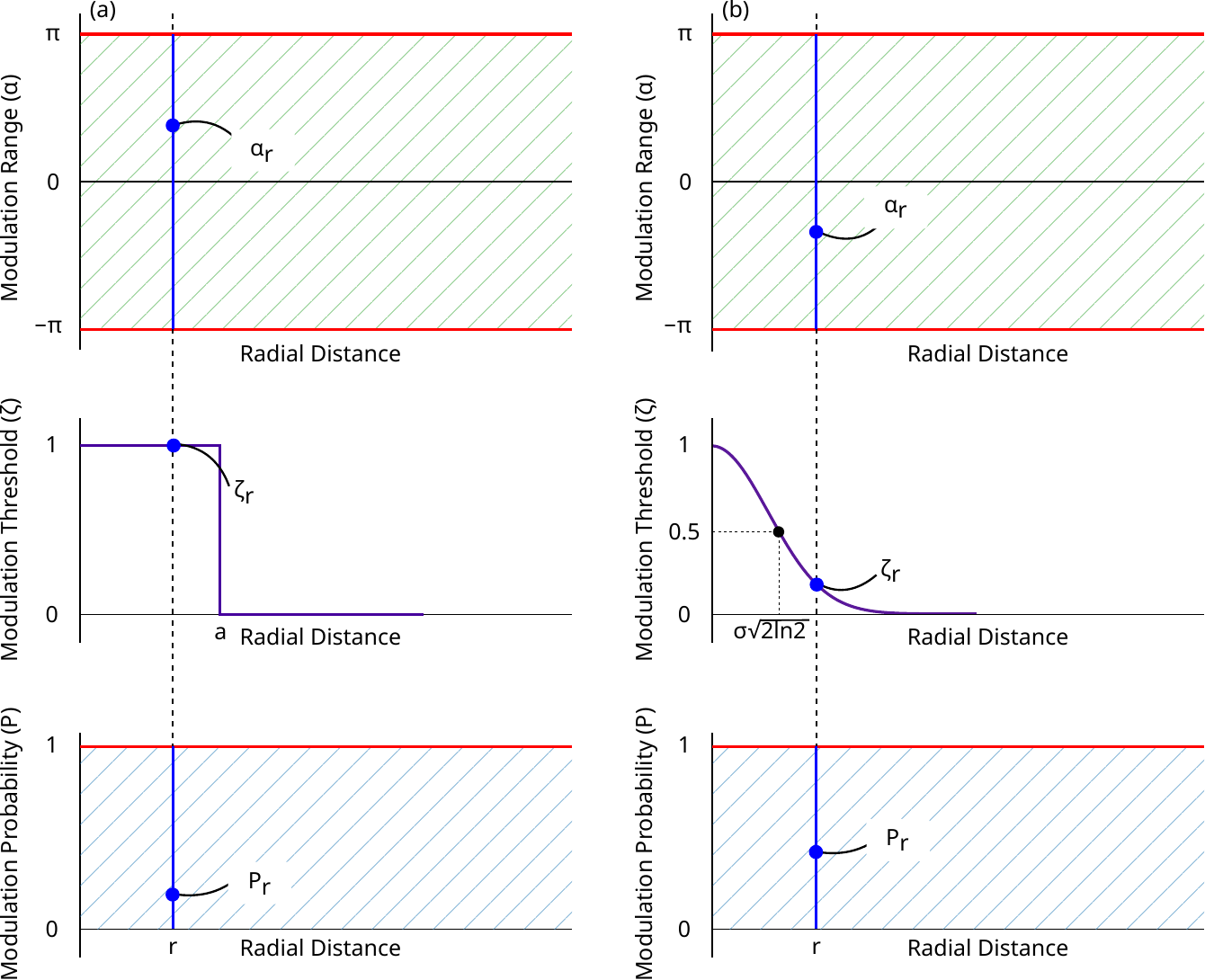}
 \caption{The mathematical models for the piece-wise modulation functions $M_\mathrm{OD}\ \mathrm{and}\ M_\mathrm{GO}$, given in Eq.~\eqref{eq:od_basic_tr} and Eq.~\eqref{eq:go_prob_mix}, are illustrated in columns (a) and (b), respectively. The first row of panels shows a uniformly distributed set of random variables within the interval $[-\pi,\pi)$, the second row shows the behaviour of the threshold corresponding to the model considered, and the third row shows a uniformly distributed set of random variables within the interval $[0,1)$. The dashed lines connecting the plots column-wise are showing an instance for a given radial distance $r$, resulting in a phase of $\alpha_r$ or $0$, depending on $p_r$ with respect to $\zeta_r$.}
\label{fig:ODGO}
\end{figure}

If we take the probability for the decoherence event to be proportional to the number of photons in that spatial point, then the decrease of the decoherence probability will trace the Gaussian shape. This can be modelled with a Gaussian probability field, where every individual point in space is compared to the respective threshold value from the probability field, and accordingly corresponding to either full or no decoherence.

This probabilistic mixture can be formulated by introducing $\beta_{\rho\phi}$, which are defined as instances of a random variable:
\begin{equation}\label{eq:manif_go}
    B_{\rho\phi} \sim \mathcal{M}(p_{\rho\phi},\zeta_{\rho},\{\{0\},A_{\rho\phi}\}) = \begin{cases}
\{0\}, & \text{if } p_{\rho\phi} \geq \zeta_{\rho} \\
A_{\rho\phi}, & \text{otherwise}
\end{cases}
\end{equation}
where $p_{\rho\phi}$ is an instance of $P_{\rho\phi} \sim \mathcal{U}_{[0,1]}$, and $\zeta_{\rho} = e^{-\rho^2/(2\sigma^2)}$ is the threshold for the probability field. This method is illustrated as a procedure in Fig. \ref{fig:ODGO}(b) where, at a given radial distance $r$, the resulting modulation phase will be randomly chosen or not depending on the probabilistic splitting. This results in the point-wise modulation function
\begin{equation}\label{eq:go_prob_mix}
    M_{\mathrm{GO}}\left(\rho ,\phi \right) = e^{i\beta _{\rho \phi }}.
\end{equation}
Notice how the resulting domain splitting does not correspond to a spatial division.

We denote this model as ``Gaussian opacity''  due to the correspondence of the random phase to an effective Gaussian opacity field. By evolving the beam with the diffraction integral given in Eq.~\eqref{eq:fresnel}, using the piece-wise modulation given in Eq.~\eqref{eq:go_prob_mix}, the azimuthal integral simplifies into $2\pi (1-\zeta_\rho)J_0({kr\rho}/{z})$, resulting in
\begin{equation}
    u_{\mathrm{GO}}\left(r,\theta \right)=\frac{k}{iz}
    \int _0^{\infty}\left(1-\zeta_\rho\right)\rho e^{i\Gamma_\rho/2}J_0\left(\frac{kr\rho}{z}\right) d\rho,
\end{equation}
which yields the intensity pattern
\begin{equation}\label{eq:go_the}
    I_\mathrm{GO}(r,z) \propto 1
    +\frac{\Gamma_\sigma^2}{1+\Gamma_\sigma^2}e^{-\Gamma_r\frac{\Gamma_\sigma}{1+\Gamma_\sigma^2}}
    -2\frac{\Gamma_\sigma^2}{1+\Gamma_\sigma^2}e^{-\frac{\Gamma_r}{2}\frac{\Gamma_\sigma}{1+\Gamma_\sigma^2}}\mathrm{c\Gamma_\sigma^{-1}s}\left(\frac{\Gamma_r}{2}\frac{1}{1+\Gamma_\sigma^2}\right),
\end{equation}
where $\mathrm{c\Upsilon s}(x)=\cos(x) + \Upsilon\sin(x)$.
The behaviour of the intensity of the central point is given by $I_\mathrm{GO}(0,z) \propto 1/(1+\Gamma_\sigma^2)$ for the case of a plane wave incident field.

\begin{figure}[h]
 \centering  
 \includegraphics[width=0.95\textwidth]{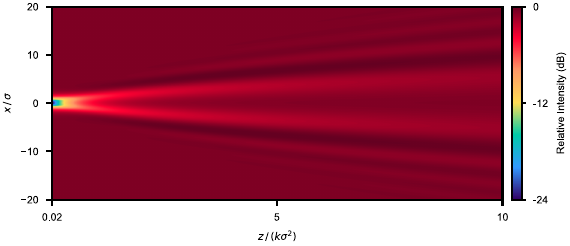}
 \caption{Theoretically expected interference pattern for the Gaussian Opacity (GO) model given in Eq.~\eqref{eq:go_the}.  The intensity pattern in the horizontal plane containing the beam axis is shown in logarithmic colour scale as a function of the distance from the spatial light modulator (SLM). The logarithmic scale is chosen for comparability.
}
\label{fig:gosim}
\end{figure}

The expected full scale intensity pattern for the GO model, containing both the near and the far field cases, is shown in Fig.~\ref{fig:gosim}. It is apparent that the interference fringes outside the geometric shadow are not present in the near field. In the far field, the interference fringes start to become similar to what is expected from the OD case, as it was shown in Fig.~\ref{fig:odsim}.

\subsection{Opacity similarity\label{app:opsim}}

We saw that the azimuthal integral for both OD and GO can be simplified into forms 
containing the shape of the probability field. For example, for the case of GO, the shape of the probability is $\zeta_\rho$, and it appears in the result of the azimuthal integral $2\pi (1-\zeta_\rho)J_0({kr\rho}/{z})$. 
In subsequent sections, we will refer to this as ``opacity similarity''.

\section{Two additional models to analyse boundary effects}\label{app:addmod}

In the previous section, we explained and elaborated the radially symmetric strong decoherence models. For the OD model, there was a circular boundary defining a hard transition in phase behaviour, whereas for the GO model the transition was smooth. This raises two questions: What happens if one has only a circular boundary (CB) with a sharp change of phase by $\pi$, and what happens if we continuously change the smoothness of this transition?
We will address these questions in the following subsections.

\subsection{Circular boundary effects}

\begin{figure}[h]
 \centering  
 \includegraphics[width=0.50\textwidth]{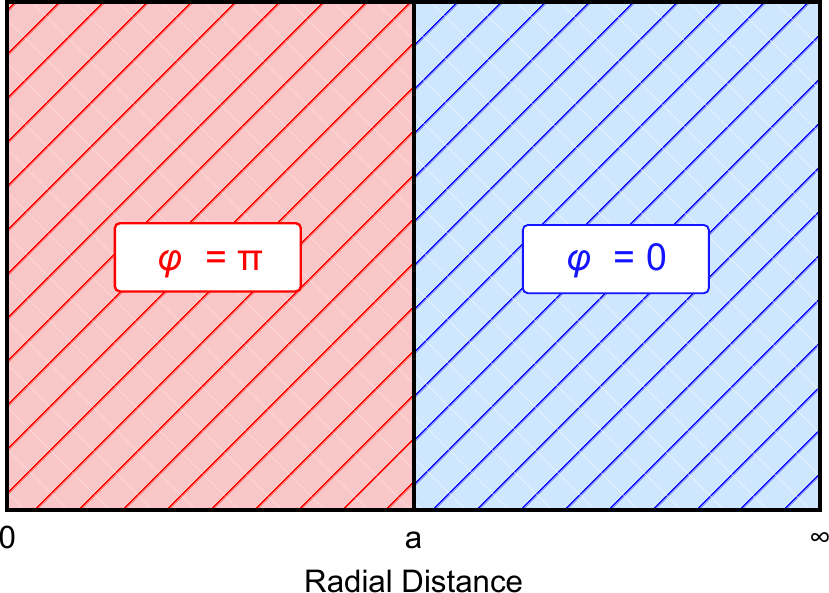}
 \caption{A basic mathematical representation of the circular boundary model, described in Eq.~\eqref{eq:cb}. The spatial region from $0$ to $a$ is assigned a constant phase of $\pi$, whereas the complementary region has no phase change.}
\label{fig:CBdom}
\end{figure}

The CB model is a simple model as illustrated in Fig. \ref{fig:CBdom}. It can be described by the following mathematical form:
\begin{equation}\label{eq:cb}
    M_{\mathrm{CB}}\left(\rho ,\phi \right)=\left\{0\le \rho <a: -1,a\le \rho:1\right\}.
\end{equation}
By evolving the beam with the diffraction integral given in Eq.~\eqref{eq:fresnel}, using the piece-wise modulation described in Eq.~\eqref{eq:cb}, the azimuthal integral simplifies to $2\pi J_o(kr\rho/z)$. The resulting field is
\begin{equation}
    u_\mathrm{CB}(r,\theta) = -\frac{k}{iz}\int_0^a\rho e^{i\Gamma_\rho/2}J_0\left(\frac{kr\rho}{z}\right) d\rho + \frac{k}{iz}\int_a^\infty\rho e^{i\Gamma_\rho/2}J_0\left(\frac{kr\rho}{z}\right) d\rho,
\end{equation}
which yields the intensity pattern
\begin{equation}\label{eq:cb_intensity}
    I_\mathrm{CB}(r,z) \propto \left|e^{-i\Gamma_r/2}+2i\frac{k}{z}\int_0^a\rho e^{i\Gamma_\rho/2}J_0\left(\frac{kr\rho}{z}\right) d\rho\right|^2.
\end{equation}
The intensity at the central point will then behave as $I_\mathrm{CB}(0,z) \propto 5-4\cos{(\Gamma_a/2)}$ for the case of a plane wave incident field.

\begin{figure}[h]
 \centering  
 \includegraphics[width=1\textwidth]{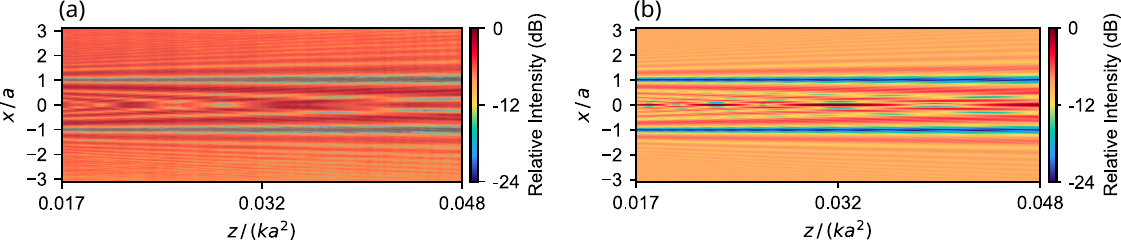}
 \caption{Interference pattern generated with the spatial light modulator (SLM) encoding the Circular Boundary (CB) model for a size parameter $a=\,800\ \mathrm{\mu m}$. Experimental results are shown in (a), while the corresponding theoretical predictions given in Eq.~\eqref{eq:cb_intensity} is shown in (b). The intensity pattern in the horizontal plane containing the beam axis is shown in logarithmic colour scale as a function of the distance from the SLM. The logarithmic scale is chosen for comparability.
 Each slice at a given $z$ is a single-pixel horizontal slice at the centre of the camera frames captured.}
\label{fig:CB}
\end{figure}

Experimental data for the case of $a = 800\ \mathrm{\mu m}$ is shown in Fig. \ref{fig:CB} (a), and the theoretical counterpart in Fig. \ref{fig:CB} (b). Notice that the experiment and the theory qualitatively agree, unlike the OD case discussed in Fig.~\ref{fig:Arago}.
For the case of OD, we argue in Sec.~\ref{ssec:sdldexp} of the main text that the SLM efficiency and the phase deviation are the reason behind the oscillatory behaviour. However, for CB, there already exists an oscillatory behaviour in the theoretical expectation. Hence, the SLM efficiency and the phase deviation cannot create an additional behavioural change, but only can cause a rescaling and translation.

\subsection{Smoothness effects}\label{app:smooth}

\begin{figure}[h!]
 \centering  
 \includegraphics[width=0.45\textwidth]{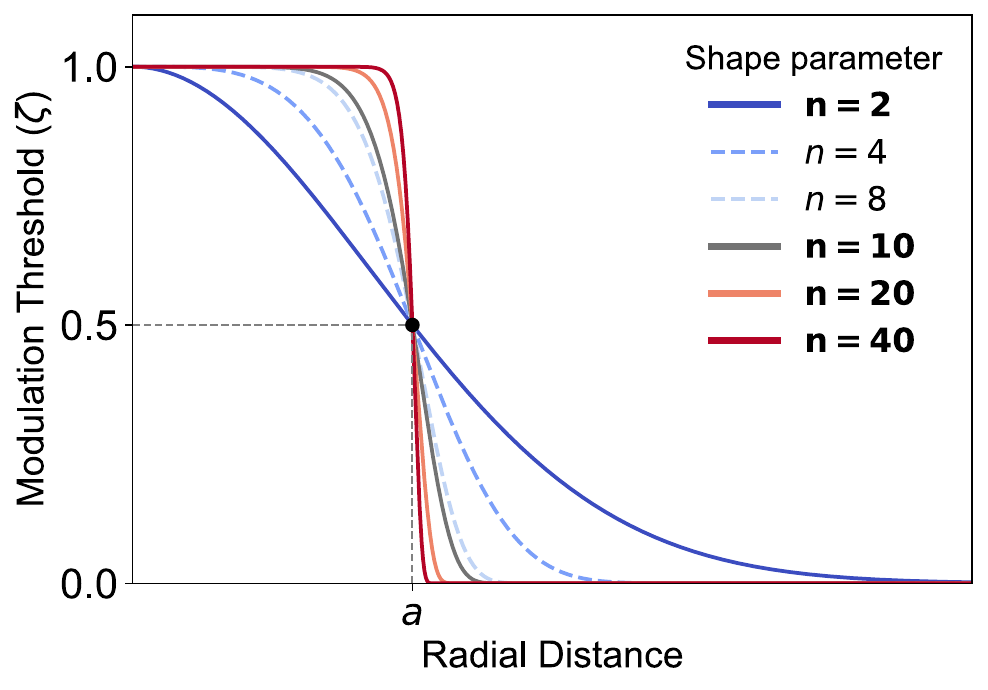}
 \caption{The threshold function in the Generalised Gaussian Opacity (GGO) model given in Eq.~\eqref{eq:manif_ggo} is shown for different values of the smoothness parameter $n$. The highlighted cases $n=\{2,10,20,40\}$ correspond to the experimentally tested cases of the GGO model.}
\label{fig:ggo_thr}
\end{figure}

The smoothness effects can be checked by generalising the GO model into the generalised Gaussian opacity (GGO) model. In particular, we need to update our formulation using a probabilistic mixture, where $\beta_{\rho\phi}$ were instances of the random variable $B_{\rho\phi}$ as described in Eq.~\eqref{eq:manif_go}. We now define $\gamma_{\rho\phi}$ as instances of a random variable $\Gamma_{\rho\phi}$:
\begin{equation}\label{eq:manif_ggo}
    \Gamma_{\rho\phi} \sim \mathcal{M}(p_{\rho\phi},\xi_{\rho},\{\{0\},A_{\rho\phi}\}) = \begin{cases}
\{0\}, & \text{if } p_{\rho\phi} \geq \xi_{\rho} \\
A_{\rho\phi}, & \text{otherwise}
\end{cases}
\end{equation}
where $\xi_{\rho} = 2^{-\left(\rho/a\right)^n}$ instead.
The model is defined for $(n\geq2)$ to ensure platykurtic fields. The resulting modulation field is
\begin{equation}\label{eq:ggo}
    M_{\mathrm{GGO}}\left(\rho ,\phi \right) =e^{i\gamma _{\rho \phi }},
\end{equation}
which, as expected from the achieved opacity similarity, maps to the corresponding generalised Gaussian opacity field. We then get the resulting field by inserting Eq.~\eqref{eq:ggo} into Eq.~\eqref{eq:fresnel}:
\begin{equation}
    u_\mathrm{GGO}(r,\theta) = \frac{k}{iz}
    \int _0^{\infty}\left(1-\xi_\rho\right)\rho e^{i\Gamma_\rho/2}J_0\left(\frac{kr\rho}{z}\right) d\rho,
\end{equation}
which results in the following intensity pattern:
\begin{equation}
    I_\mathrm{GGO}(r,z) \propto \left|e^{-i\Gamma_r/2}+i\frac{k}{z}\int _0^{\infty}\xi_\rho\rho e^{i\Gamma_\rho/2} J_0\left(\frac{kr\rho}{z}\right) d\rho\right|^2.
\end{equation} 
Notice that, as $n$ increases, $\lim_{n\to\infty}I_\mathrm{GGO}(r,z) = I_\mathrm{OD}(r,z)$, the GGO model approaches OD. Fig.~\ref{fig:ggo_thr} depicts how the smoothness changes for different cases of $n$, while keeping the half-point $a$ constant.

\begin{figure}[h!]
 \centering  
 \includegraphics[width=1\textwidth]{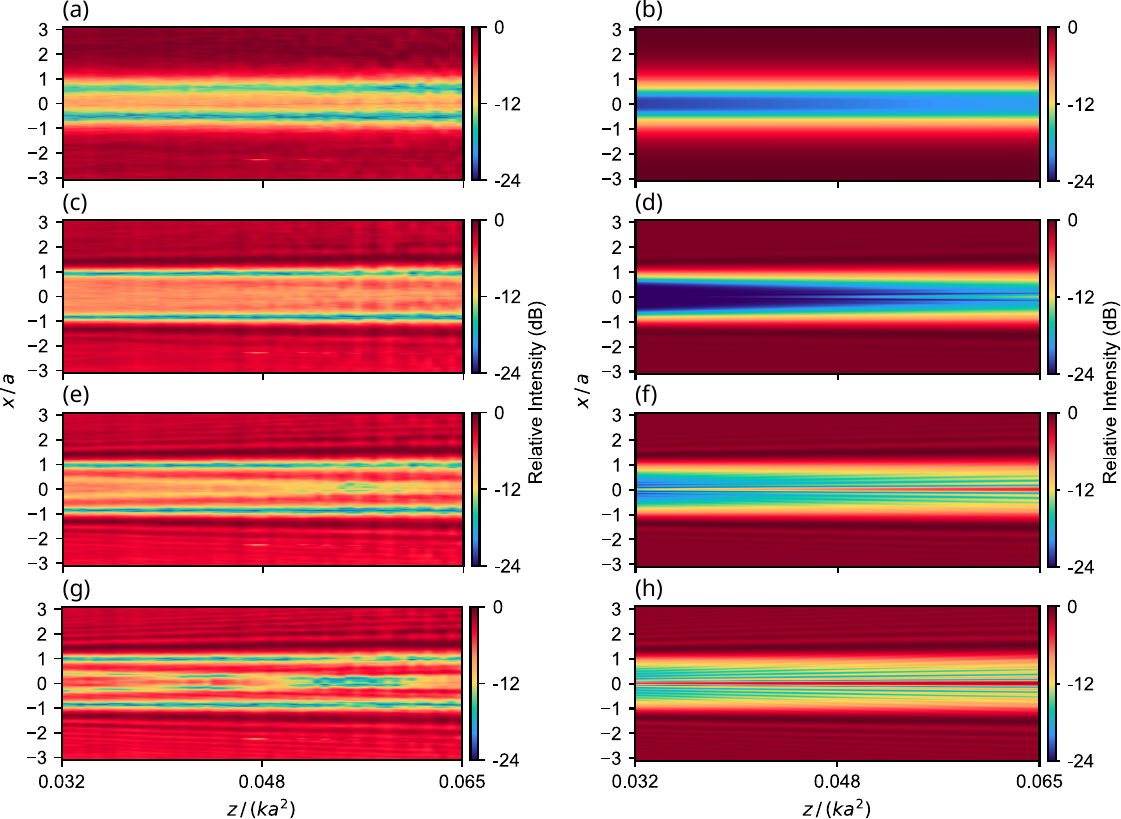}
 \caption{Interference pattern generated with the spatial light modulator (SLM) encoding the Generalized Gaussian Opacity (GGO) model for a size parameter $a=\,800\ \mathrm{\mu m}$. Experimental results and the corresponding theoretical predictions are shown row-wise for different smoothness parameters (a-b) $n=2$, (c-d) $n=10$, (e-f) $n=20$, (g-h) $n=40$. The intensity pattern in the horizontal plane containing the beam axis is shown in logarithmic color scale for different distances from the SLM. The logarithmic scale is chosen for comparability. Each slice at a given $z$ is a single-pixel horizontal slice at the centre of the camera images for the experimental data.}
\label{fig:ggo_transition}
\end{figure}

The experimental data and the corresponding theoretical prediction for $a = 800\ \mathrm{\mu m}$ and $n = 2, 10, 20, \mathrm{and}\ 40$, the highlighted cases in Fig. \ref{fig:ggo_thr}, are provided in Fig. \ref{fig:ggo_transition}. Note that the interference fringes begin to appear around the geometric shadow as $n$ increases. This behaviour is apparent for both the experiments and the theory. This confirms our conclusion that the hard boundary causes diffractional fringes in the near field. Also, the Arago spot starts to form as the smoothness parameter increases.

\section{Modelling of periodic decoherence}\label{app:po}
For the periodic domain local decoherence, we use an accordion-like optical lattice formed along the vertical axis as horizontal stripes, following the spatial profile:
\begin{equation}
    \varpi_{\eta} = e^{-\frac{2\eta^{2}\cos^{2}(\varphi/2)}{w^{2}}}\cos^{2}\left(2\pi \eta\frac{\sin(\varphi/2)}{l}\right),
\end{equation}
where $\eta$ is the coordinate along the vertical axis; $w$, $l$ and $\varphi$ are the model parameters. These parameters can be mapped to two phenomenological dependent variables, namely the effective number of stripes $n$ and the grid period $\varsigma$. The latter is easy to calculate by checking the periodicity of the equation, resulting in $\varsigma=l/2\sin(\varphi/2)$. However, the effective grid number requires proper derivation.

Let us define a cut-off $K$ to decide which peaks will be counted as a part of the created dark grid. Here, the dark grid refers to the part of the light that will get fully dephased with high probability.
Using integer multiples of the grid period $\eta_j = j\varsigma$, we can use condition $\varpi_{\eta_j} = K$ to see where the cut-off line is tangential to the first peak that will not be taken into account. Considering that the $(j-1)$th peak is the last peak to be taken into account, the total dark grid number will be $2(j-1)+1$. Hence, the total light grid, which is the complementary of the dark grid introduced earlier, will consist of $2j$ stripes, therefore the resulting effective number of stripes is $n=2(w/l)\sqrt{-2\ln(K)}\tan(\varphi/2)$. In the periodic opacity (PO) description given in Sec.~\ref{sec:corresp} of the main text, we chose $K=1/2$.

Since the effective number of stripes must be the same for the optics and the matter wave counterparts, we have $w/l=w_\mathrm{dec}/\lambda_\mathrm{dec}$. Similarly, from the equivalence of the angle $\varphi$ between for these two cases, we have $\varsigma/l=\varsigma_\mathrm{dec}/\lambda_\mathrm{dec}$. Using the correspondence equation Eq.~\eqref{eq:plcorr} given in Sec.~\ref{sec:corresp} of the main text, the optical Talbot length $z_\mathrm{T}=2\varsigma^2/\lambda$ and the matter-wave Talbot time $t_\mathrm{T}=2m\varsigma_\mathrm{dec}^2/h$, one can derive the following size scaling:
\begin{equation}
    \frac{r}{\rho} = \frac{\varsigma}{\varsigma_\mathrm{dec}} = \frac{l}{\lambda_\mathrm{dec}} = \frac{w}{w_\mathrm{dec}} = \sqrt{\frac{z_\mathrm{T}}{t_\mathrm{T}}\frac{m\lambda}{h}}.
\end{equation}

The accordion-like optical lattice can easily be turned into a decoherence distribution model similar to GO and GGO by using a probabilistic mixture formulation with $\delta_{\xi\eta}$ as instances of the random variable $\Delta_{\xi\eta}$:
\begin{equation}
    \Delta_{\xi\eta} \sim \mathcal{M}(p_{\xi\eta},\varpi_\eta,\{\{0\},A_{\xi\eta}\}) = \begin{cases}
\{0\}, & \text{if } p_{\xi\eta} \geq \varpi_{\eta} \\
A_{\xi\eta}, & \text{otherwise}
\end{cases}
\end{equation}
where $A_{\xi\eta}\sim \mathcal{U}_{\left[-\pi ,\pi \right)}$.
The resulting modulation field is
\begin{equation}\label{eq:po}
    M_\mathrm{PO}(\xi,\eta)=e^{i\delta_{\xi\eta}}.
\end{equation}
By evolving the beam in the Cartesian version of Eq.~\eqref{eq:fresnel} with this point-wise modulation Eq.~\eqref{eq:po}, we get the evaluated field:
\begin{equation}
    u_\mathrm{PO}(x,y) = \sqrt{\frac{k}{2\pi i z}}e^{-i\Gamma_x/2}\int _{-\infty}^{\infty}(1-\varpi_\eta)e^{i\Gamma_\eta/2}e^{-i\frac{ky\eta}{z}} d\eta, 
\end{equation}
from the opacity similarity, which results in the following intensity pattern:
\begin{equation}
    I_\mathrm{PO}(y,z) \propto \left|e^{-i\Gamma_y/2} - \sqrt{\frac{k}{2\pi i z}}\int _{-\infty}^{\infty}\varpi_\eta e^{i\Gamma_\eta/2}e^{-i\frac{ky\eta}{z}} d\eta\right|^2.
\end{equation}

\section{Modelling of SLM to explain observed central point intensity behaviour}\label{app:fits}

The models discussed in Appendix~\ref{app:models} are for an ideal case, where the modulation acts on an infinite plane-wave with perfect phase modulation. Such perfect phase modulation is impossible to achieve in an experiment. The spatial light modulator (SLM) we use to implement the dephasing has efficiency $\eta$, which means that the incoming beam is split into two coherent parts, where one part experiences the modulation, while the other part is simply reflected with a phase shift $\pi+\varepsilon$. If we take into account the efficiency $\eta$ and a deviation $\varepsilon$ from a perfect $\pi$ phase shift upon reflection, the corresponding behaviour of the central intensities for OD and GO, respectively, will be:
\begin{equation}\label{eq:fit_slm}
\begin{aligned}
    I_{\mathrm{OD}}'\left(0,z\right)&\propto 1-2\sqrt{\eta(1-\eta)}\cos\left(\Gamma_a/2-\varepsilon\right) \\
    I_{\mathrm{GO}}'\left(0,z\right)&\propto 1-\eta\frac{\Gamma_\sigma^2}{1+\Gamma_\sigma^2}-2\sqrt{\eta(1-\eta)}\frac{1}{1+\Gamma_\sigma^2}\mathrm{c\Gamma_\sigma s}(\varepsilon)
\end{aligned}
\end{equation}

\begin{figure}[h]
\centering
\begin{minipage}{\textwidth}
\centering
\adjustbox{valign=m}{\includegraphics[width=0.37\textwidth]{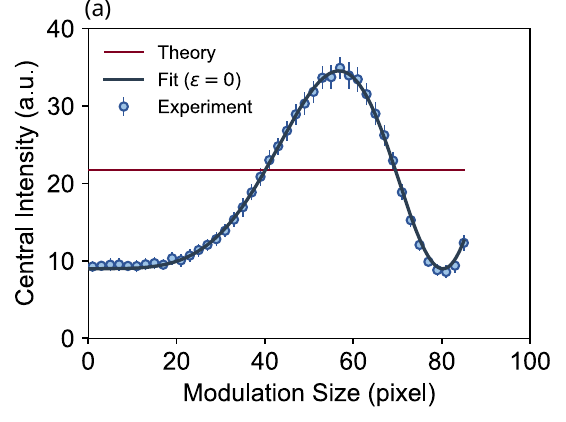}}
\hspace{1.1cm} 
\adjustbox{valign=m}{
    \begin{tabular}{c|c}
        Fit & OD\\
        \hline
        0 & $\varepsilon$ (rad)\\
        0.90 & $\eta$\\
        21.30 & $B$\\
        0.44 & $C$\\
        0.77 & $K$ (mm)\\
        0.99 & $R^2$
    \end{tabular}
}
\end{minipage}

\vspace{0.5em}
\begin{minipage}{\textwidth}
\centering
\adjustbox{valign=m}{\includegraphics[width=0.37\textwidth]{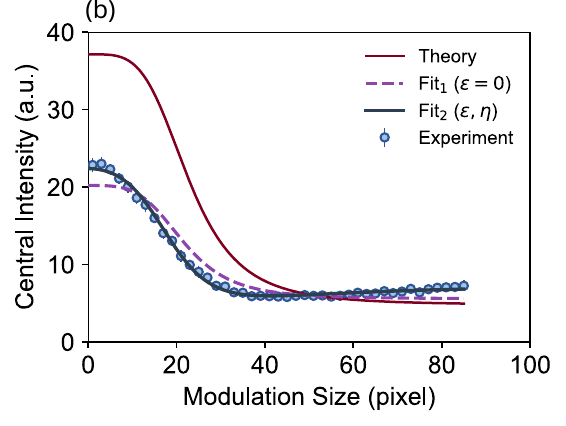}}
\hspace{0.20cm}
\adjustbox{valign=m}{
    \begin{tabular}{cc|c}
        $\mathrm{Fit}_2$ & $\mathrm{Fit}_1$ & GO\\
        \hline
        0.60 & 0 & $\varepsilon$ (rad)\\
        0.92 & 0.99 & $\eta$\\
        32.29 & 14.62 & $B$\\
        4.79 & 5.59 & $C$\\
        0.77 & 0.80 & $K$ (mm)\\
        0.99 & 0.94 & $R^2$
    \end{tabular}
}
\end{minipage}

\caption{Intensity of light in the centre of the interference pattern generated with the spatial light modulator (SLM) encoding (a) the Opaque Disk (OD) point-wise modulation and (b) the Gaussian Opacity (GO) point-wise modulation, for varying sizes of the modulated region.
The exposure times were selected to be 400 ms for OD, and 1000 ms for GO. The number of frames taken were selected to be 50 for OD, and 20 for GO.
The distance from the SLM is {$z=40$ cm}. For both OD and GO, the points represent the measured data with the corresponding error bars indicating the standard deviations. For OD, the solid black line is the corresponding fitted function~\eqref{eq:fit_slm}. For GO, the dashed line is the corresponding fitted function~\eqref{eq:fit_slm} without the phase deviation $\varepsilon$, and the black line is the fit with the full model. For both OD and GO, the red lines are the resulting theoretical expectations.}
\label{fig:AragoCenter}

\end{figure}

As an experimental quantification for the observed deviations, here we show the intensity at the centre of the interference pattern as a function of the modulation size for both the OD model in Fig.~\ref{fig:AragoCenter}(a) and the GO model in Fig.~\ref{fig:AragoCenter}(b). Each experimental data point was obtained as the maximum intensity on a grid of 1x7 pixels in the centre of the beam to reduce the impact of the finite size of the rectangular aperture of the active area of the SLM. 
The fit curves are obtained by using Eq. {\eqref{eq:fit_slm}}, in the form $BI'(0,z)+ C$, to accommodate the laser intensity seen by the camera ($B$), and possible constant background in the experimental setup ($C$).
The parameter $\Gamma_j=kj^2/z$ can be rewritten as $\Gamma_j=Kj_p^2/z_\mathrm{mm}$, incorporating $K=10^3kp^2\doteq0.77$, where $p$ is the pixel size, $j_p$ is the radial element $j$ in pixels, and $z_\mathrm{mm}$ is the distance $z$ in millimetres.

The tables presented in Fig.~\ref{fig:AragoCenter} contain the fit parameters for the corresponding fits illustrated in (a) and (b). Note that in both cases (for OD and GO), the goodness of the fit ($R^2$) is reasonably high. However, there are discrepancies between the sets of fit parameters $\{B, C, K,\eta,\varepsilon\}$. This can be understood as follows: The mathematical formulations of the fit models given in Eq.~\eqref{eq:fit_slm} 
do not necessarily represent the full description of the real phenomena. The parts within the formulations are indeed in good alignment with the underlying physics. The terms $\cos({\Gamma_a/2})$, $1/(1+\Gamma_\sigma^2)$, $\Gamma_\sigma/(1+\Gamma_\sigma^2)$, and $\Gamma_\sigma^2/(1+\Gamma_\sigma^2)$ in the fit functions correspond to terms occurring in the theoretical expectations for the OD, GO, and CB models. Since $\Gamma_j$ is a function of the fit parameter $K$, the latter is a fundamental element required to be in a fit. Introducing the parameter $\eta$ in addition to $K$ already allows one to achieve a reasonably good fit for the OD case. Therefore, one could argue that $\eta$ is a physical parameter. In fact, descriptions and observations of this ``diffraction efficiency'' already exist in the literature~\cite{Buralli1992,Moreno2011}. However, one can see that the fits for the OD and GO models do not agree on the value of the parameter $\varepsilon$, although they agree on the value of $K$ and $\eta$. Possible additional refinements of our fit models are beyond the scope of this paper.

\section{Smoothening mask}\label{app:mask}
Since our experimental setup has a collimated beam with a waist larger than the SLM, diffraction at the rectangular aperture would lead to horizontal and vertical fringes.
This arises from the rectangular aperture of the SLM acting as a hard boundary. 
To suppress these unwanted effects, we introduced a ``smoothening mask''  close to the edge of the SLM screen.

\begin{figure}[h!]
\centering 
\includegraphics[width=0.5\textwidth]{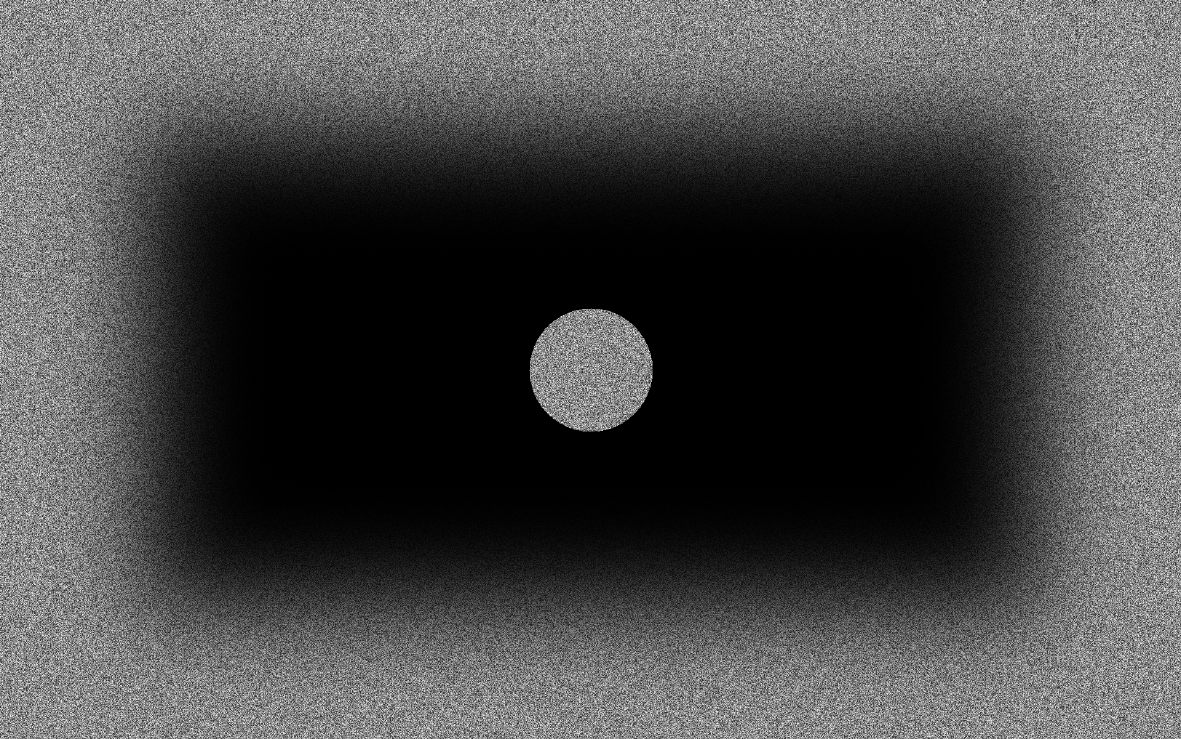}
\caption{The full image of the pattern for the opaque disk (OD) model, generated by the spatial light modulator (SLM). The central disk shaped region is where the Eq.~\eqref{eq:od_basic_tr} is evaluated. The outer rectangular region on the edges is the smoothening mask.}
\label{fig:mask}
\end{figure}

The smoothening mask, together with the OD model operated at the centre, is shown in Fig.~\ref{fig:mask}. The idea behind the mask follows the mathematically established opacity similarity in Appendix ~\ref{app:opsim}, and the boundary smoothening effects described in Appendix ~\ref{app:smooth}.

\bibliographystyle{apsrev4-2}
\bibliography{bibliography_list}

\begin{thebibliography}{28}%
\makeatletter
\providecommand \@ifxundefined [1]{%
 \@ifx{#1\undefined}
}%
\providecommand \@ifnum [1]{%
 \ifnum #1\expandafter \@firstoftwo
 \else \expandafter \@secondoftwo
 \fi
}%
\providecommand \@ifx [1]{%
 \ifx #1\expandafter \@firstoftwo
 \else \expandafter \@secondoftwo
 \fi
}%
\providecommand \natexlab [1]{#1}%
\providecommand \enquote  [1]{``#1''}%
\providecommand \bibnamefont  [1]{#1}%
\providecommand \bibfnamefont [1]{#1}%
\providecommand \citenamefont [1]{#1}%
\providecommand \href@noop [0]{\@secondoftwo}%
\providecommand \href [0]{\begingroup \@sanitize@url \@href}%
\providecommand \@href[1]{\@@startlink{#1}\@@href}%
\providecommand \@@href[1]{\endgroup#1\@@endlink}%
\providecommand \@sanitize@url [0]{\catcode `\\12\catcode `\$12\catcode
  `\&12\catcode `\#12\catcode `\^12\catcode `\_12\catcode `\%12\relax}%
\providecommand \@@startlink[1]{}%
\providecommand \@@endlink[0]{}%
\providecommand \url  [0]{\begingroup\@sanitize@url \@url }%
\providecommand \@url [1]{\endgroup\@href {#1}{\urlprefix }}%
\providecommand \urlprefix  [0]{URL }%
\providecommand \Eprint [0]{\href }%
\providecommand \doibase [0]{https://doi.org/}%
\providecommand \selectlanguage [0]{\@gobble}%
\providecommand \bibinfo  [0]{\@secondoftwo}%
\providecommand \bibfield  [0]{\@secondoftwo}%
\providecommand \translation [1]{[#1]}%
\providecommand \BibitemOpen [0]{}%
\providecommand \bibitemStop [0]{}%
\providecommand \bibitemNoStop [0]{.\EOS\space}%
\providecommand \EOS [0]{\spacefactor3000\relax}%
\providecommand \BibitemShut  [1]{\csname bibitem#1\endcsname}%
\let\auto@bib@innerbib\@empty
\bibitem [{\citenamefont {Joos}\ and\ \citenamefont {Zeh}(1985)}]{Joos1985a}%
  \BibitemOpen
  \bibfield  {author} {\bibinfo {author} {\bibfnamefont {E.}~\bibnamefont
  {Joos}}\ and\ \bibinfo {author} {\bibfnamefont {H.~D.}\ \bibnamefont {Zeh}},\
  }\href {http://dx.doi.org/10.1007/BF01725541} {\bibfield  {journal} {\bibinfo
   {journal} {Zeitschrift für Physik B Condensed Matter}\ }\textbf {\bibinfo
  {volume} {59}},\ \bibinfo {pages} {223} (\bibinfo {year} {1985})},\ \bibinfo
  {note} {10.1007/BF01725541}\BibitemShut {NoStop}%
\bibitem [{\citenamefont {Zurek}(2003)}]{Zurek2003a}%
  \BibitemOpen
  \bibfield  {author} {\bibinfo {author} {\bibfnamefont {W.~H.}\ \bibnamefont
  {Zurek}},\ }\href {https://doi.org/10.1103/RevModPhys.75.715} {\bibfield
  {journal} {\bibinfo  {journal} {Rev. Mod. Phys.}\ }\textbf {\bibinfo {volume}
  {75}},\ \bibinfo {pages} {715} (\bibinfo {year} {2003})}\BibitemShut
  {NoStop}%
\bibitem [{\citenamefont {Schlosshauer}(2019)}]{Schlosshauer2019a}%
  \BibitemOpen
  \bibfield  {author} {\bibinfo {author} {\bibfnamefont {M.}~\bibnamefont
  {Schlosshauer}},\ }\href {https://doi.org/10.1016/j.physrep.2019.10.001}
  {\bibfield  {journal} {\bibinfo  {journal} {Physics Reports}\ }\bibinfo
  {series} {Quantum decoherence},\ \textbf {\bibinfo {volume} {831}},\ \bibinfo
  {pages} {1} (\bibinfo {year} {2019})}\BibitemShut {NoStop}%
\bibitem [{\citenamefont {Hornberger}\ \emph {et~al.}(2003)\citenamefont
  {Hornberger}, \citenamefont {Uttenthaler}, \citenamefont {Brezger},
  \citenamefont {Hackerm\"uller}, \citenamefont {Arndt},\ and\ \citenamefont
  {Zeilinger}}]{Hornberger2003b}%
  \BibitemOpen
  \bibfield  {author} {\bibinfo {author} {\bibfnamefont {K.}~\bibnamefont
  {Hornberger}}, \bibinfo {author} {\bibfnamefont {S.}~\bibnamefont
  {Uttenthaler}}, \bibinfo {author} {\bibfnamefont {B.}~\bibnamefont
  {Brezger}}, \bibinfo {author} {\bibfnamefont {L.}~\bibnamefont
  {Hackerm\"uller}}, \bibinfo {author} {\bibfnamefont {M.}~\bibnamefont
  {Arndt}},\ and\ \bibinfo {author} {\bibfnamefont {A.}~\bibnamefont
  {Zeilinger}},\ }\href {https://doi.org/10.1103/PhysRevLett.90.160401}
  {\bibfield  {journal} {\bibinfo  {journal} {Phys. Rev. Lett.}\ }\textbf
  {\bibinfo {volume} {90}},\ \bibinfo {pages} {160401} (\bibinfo {year}
  {2003})}\BibitemShut {NoStop}%
\bibitem [{\citenamefont {Hackerm\"{u}ller}\ \emph {et~al.}(2004)\citenamefont
  {Hackerm\"{u}ller}, \citenamefont {Hornberger}, \citenamefont {Brezger},
  \citenamefont {Zeilinger},\ and\ \citenamefont {Arndt}}]{Hackermueller2004a}%
  \BibitemOpen
  \bibfield  {author} {\bibinfo {author} {\bibfnamefont {L.}~\bibnamefont
  {Hackerm\"{u}ller}}, \bibinfo {author} {\bibfnamefont {K.}~\bibnamefont
  {Hornberger}}, \bibinfo {author} {\bibfnamefont {B.}~\bibnamefont {Brezger}},
  \bibinfo {author} {\bibfnamefont {A.}~\bibnamefont {Zeilinger}},\ and\
  \bibinfo {author} {\bibfnamefont {M.}~\bibnamefont {Arndt}},\ }\href
  {https://doi.org/10.1038/nature02276} {\bibfield  {journal} {\bibinfo
  {journal} {Nature}\ }\textbf {\bibinfo {volume} {427}},\ \bibinfo {pages}
  {711–714} (\bibinfo {year} {2004})}\BibitemShut {NoStop}%
\bibitem [{\citenamefont {Gerlich}\ \emph {et~al.}(2007)\citenamefont
  {Gerlich}, \citenamefont {Hackerm\"{u}ller}, \citenamefont {Hornberger},
  \citenamefont {Stibor}, \citenamefont {Ulbricht}, \citenamefont {Gring},
  \citenamefont {Goldfarb}, \citenamefont {Savas}, \citenamefont {M\"{u}ri},
  \citenamefont {Mayor},\ and\ \citenamefont {Arndt}}]{Gerlich2007}%
  \BibitemOpen
  \bibfield  {author} {\bibinfo {author} {\bibfnamefont {S.}~\bibnamefont
  {Gerlich}}, \bibinfo {author} {\bibfnamefont {L.}~\bibnamefont
  {Hackerm\"{u}ller}}, \bibinfo {author} {\bibfnamefont {K.}~\bibnamefont
  {Hornberger}}, \bibinfo {author} {\bibfnamefont {A.}~\bibnamefont {Stibor}},
  \bibinfo {author} {\bibfnamefont {H.}~\bibnamefont {Ulbricht}}, \bibinfo
  {author} {\bibfnamefont {M.}~\bibnamefont {Gring}}, \bibinfo {author}
  {\bibfnamefont {F.}~\bibnamefont {Goldfarb}}, \bibinfo {author}
  {\bibfnamefont {T.}~\bibnamefont {Savas}}, \bibinfo {author} {\bibfnamefont
  {M.}~\bibnamefont {M\"{u}ri}}, \bibinfo {author} {\bibfnamefont
  {M.}~\bibnamefont {Mayor}},\ and\ \bibinfo {author} {\bibfnamefont
  {M.}~\bibnamefont {Arndt}},\ }\href {https://doi.org/10.1038/nphys701}
  {\bibfield  {journal} {\bibinfo  {journal} {Nature Physics}\ }\textbf
  {\bibinfo {volume} {3}},\ \bibinfo {pages} {711–715} (\bibinfo {year}
  {2007})}\BibitemShut {NoStop}%
\bibitem [{\citenamefont {Hornberger}\ \emph {et~al.}(2009)\citenamefont
  {Hornberger}, \citenamefont {Gerlich}, \citenamefont {Ulbricht},
  \citenamefont {Hackerm{\"{u}}ller}, \citenamefont {Nimmrichter},
  \citenamefont {{V Goldt}}, \citenamefont {Boltalina},\ and\ \citenamefont
  {Arndt}}]{Hornberger2009a}%
  \BibitemOpen
  \bibfield  {author} {\bibinfo {author} {\bibfnamefont {K.}~\bibnamefont
  {Hornberger}}, \bibinfo {author} {\bibfnamefont {S.}~\bibnamefont {Gerlich}},
  \bibinfo {author} {\bibfnamefont {H.}~\bibnamefont {Ulbricht}}, \bibinfo
  {author} {\bibfnamefont {L.}~\bibnamefont {Hackerm{\"{u}}ller}}, \bibinfo
  {author} {\bibfnamefont {S.}~\bibnamefont {Nimmrichter}}, \bibinfo {author}
  {\bibfnamefont {I.}~\bibnamefont {{V Goldt}}}, \bibinfo {author}
  {\bibfnamefont {O.}~\bibnamefont {Boltalina}},\ and\ \bibinfo {author}
  {\bibfnamefont {M.}~\bibnamefont {Arndt}},\ }\href
  {https://doi.org/10.1088/1367-2630/11/4/043032} {\bibfield  {journal}
  {\bibinfo  {journal} {New Journal of Physics}\ }\textbf {\bibinfo {volume}
  {11}},\ \bibinfo {pages} {043032} (\bibinfo {year} {2009})}\BibitemShut
  {NoStop}%
\bibitem [{\citenamefont {Nimmrichter}\ \emph {et~al.}(2011)\citenamefont
  {Nimmrichter}, \citenamefont {Haslinger}, \citenamefont {Hornberger},\ and\
  \citenamefont {Arndt}}]{Nimmrichter2011a}%
  \BibitemOpen
  \bibfield  {author} {\bibinfo {author} {\bibfnamefont {S.}~\bibnamefont
  {Nimmrichter}}, \bibinfo {author} {\bibfnamefont {P.}~\bibnamefont
  {Haslinger}}, \bibinfo {author} {\bibfnamefont {K.}~\bibnamefont
  {Hornberger}},\ and\ \bibinfo {author} {\bibfnamefont {M.}~\bibnamefont
  {Arndt}},\ }\href {https://doi.org/10.1088/1367-2630/13/7/075002} {\bibfield
  {journal} {\bibinfo  {journal} {New Journal of Physics}\ }\textbf {\bibinfo
  {volume} {13}},\ \bibinfo {pages} {075002} (\bibinfo {year}
  {2011})}\BibitemShut {NoStop}%
\bibitem [{\citenamefont {Haslinger}\ \emph {et~al.}(2013)\citenamefont
  {Haslinger}, \citenamefont {D{\"o}rre}, \citenamefont {Geyer}, \citenamefont
  {Rodewald}, \citenamefont {Nimmrichter},\ and\ \citenamefont
  {Arndt}}]{Haslinger2013a}%
  \BibitemOpen
  \bibfield  {author} {\bibinfo {author} {\bibfnamefont {P.}~\bibnamefont
  {Haslinger}}, \bibinfo {author} {\bibfnamefont {N.}~\bibnamefont
  {D{\"o}rre}}, \bibinfo {author} {\bibfnamefont {P.}~\bibnamefont {Geyer}},
  \bibinfo {author} {\bibfnamefont {J.}~\bibnamefont {Rodewald}}, \bibinfo
  {author} {\bibfnamefont {S.}~\bibnamefont {Nimmrichter}},\ and\ \bibinfo
  {author} {\bibfnamefont {M.}~\bibnamefont {Arndt}},\ }\href
  {https://doi.org/10.1038/nphys2542} {\bibfield  {journal} {\bibinfo
  {journal} {Nat. Phys.}\ }\textbf {\bibinfo {volume} {9}},\ \bibinfo {pages}
  {144} (\bibinfo {year} {2013})}\BibitemShut {NoStop}%
\bibitem [{\citenamefont {Eibenberger}\ \emph {et~al.}(2013)\citenamefont
  {Eibenberger}, \citenamefont {Gerlich}, \citenamefont {Arndt}, \citenamefont
  {Mayor},\ and\ \citenamefont {T\"{u}xen}}]{Eibenberger2013a}%
  \BibitemOpen
  \bibfield  {author} {\bibinfo {author} {\bibfnamefont {S.}~\bibnamefont
  {Eibenberger}}, \bibinfo {author} {\bibfnamefont {S.}~\bibnamefont
  {Gerlich}}, \bibinfo {author} {\bibfnamefont {M.}~\bibnamefont {Arndt}},
  \bibinfo {author} {\bibfnamefont {M.}~\bibnamefont {Mayor}},\ and\ \bibinfo
  {author} {\bibfnamefont {J.}~\bibnamefont {T\"{u}xen}},\ }\href
  {https://doi.org/10.1039/c3cp51500a} {\bibfield  {journal} {\bibinfo
  {journal} {Physical Chemistry Chemical Physics}\ }\textbf {\bibinfo {volume}
  {15}},\ \bibinfo {pages} {14696–14700} (\bibinfo {year}
  {2013})}\BibitemShut {NoStop}%
\bibitem [{\citenamefont {Fein}\ \emph {et~al.}(2019)\citenamefont {Fein},
  \citenamefont {Geyer}, \citenamefont {Zwick}, \citenamefont {Kia{\l}ka},
  \citenamefont {Pedalino}, \citenamefont {Mayor}, \citenamefont {Gerlich},\
  and\ \citenamefont {Arndt}}]{Fein2019a}%
  \BibitemOpen
  \bibfield  {author} {\bibinfo {author} {\bibfnamefont {Y.~Y.}\ \bibnamefont
  {Fein}}, \bibinfo {author} {\bibfnamefont {P.}~\bibnamefont {Geyer}},
  \bibinfo {author} {\bibfnamefont {P.}~\bibnamefont {Zwick}}, \bibinfo
  {author} {\bibfnamefont {F.}~\bibnamefont {Kia{\l}ka}}, \bibinfo {author}
  {\bibfnamefont {S.}~\bibnamefont {Pedalino}}, \bibinfo {author}
  {\bibfnamefont {M.}~\bibnamefont {Mayor}}, \bibinfo {author} {\bibfnamefont
  {S.}~\bibnamefont {Gerlich}},\ and\ \bibinfo {author} {\bibfnamefont
  {M.}~\bibnamefont {Arndt}},\ }\href
  {https://doi.org/10.1038/s41567-019-0663-9} {\bibfield  {journal} {\bibinfo
  {journal} {Nature Physics}\ ,\ \bibinfo {pages} {1}} (\bibinfo {year}
  {2019})}\BibitemShut {NoStop}%
\bibitem [{\citenamefont {Pedalino}\ \emph {et~al.}(2026)\citenamefont
  {Pedalino}, \citenamefont {Ramírez-Galindo}, \citenamefont {Ferstl},
  \citenamefont {Hornberger}, \citenamefont {Arndt},\ and\ \citenamefont
  {Gerlich}}]{Pedalino2026a}%
  \BibitemOpen
  \bibfield  {author} {\bibinfo {author} {\bibfnamefont {S.}~\bibnamefont
  {Pedalino}}, \bibinfo {author} {\bibfnamefont {B.~E.}\ \bibnamefont
  {Ramírez-Galindo}}, \bibinfo {author} {\bibfnamefont {R.}~\bibnamefont
  {Ferstl}}, \bibinfo {author} {\bibfnamefont {K.}~\bibnamefont {Hornberger}},
  \bibinfo {author} {\bibfnamefont {M.}~\bibnamefont {Arndt}},\ and\ \bibinfo
  {author} {\bibfnamefont {S.}~\bibnamefont {Gerlich}},\ }\href
  {https://doi.org/10.1038/s41586-025-09917-9} {\bibfield  {journal} {\bibinfo
  {journal} {Nature}\ }\textbf {\bibinfo {volume} {649}},\ \bibinfo {pages}
  {866–870} (\bibinfo {year} {2026})}\BibitemShut {NoStop}%
\bibitem [{\citenamefont {Brezger}\ \emph {et~al.}(2002)\citenamefont
  {Brezger}, \citenamefont {Hackerm\"uller}, \citenamefont {Uttenthaler},
  \citenamefont {Petschinka}, \citenamefont {Arndt},\ and\ \citenamefont
  {Zeilinger}}]{Brezger2002a}%
  \BibitemOpen
  \bibfield  {author} {\bibinfo {author} {\bibfnamefont {B.}~\bibnamefont
  {Brezger}}, \bibinfo {author} {\bibfnamefont {L.}~\bibnamefont
  {Hackerm\"uller}}, \bibinfo {author} {\bibfnamefont {S.}~\bibnamefont
  {Uttenthaler}}, \bibinfo {author} {\bibfnamefont {J.}~\bibnamefont
  {Petschinka}}, \bibinfo {author} {\bibfnamefont {M.}~\bibnamefont {Arndt}},\
  and\ \bibinfo {author} {\bibfnamefont {A.}~\bibnamefont {Zeilinger}},\ }\href
  {https://doi.org/10.1103/PhysRevLett.88.100404} {\bibfield  {journal}
  {\bibinfo  {journal} {Phys. Rev. Lett.}\ }\textbf {\bibinfo {volume} {88}},\
  \bibinfo {pages} {100404} (\bibinfo {year} {2002})}\BibitemShut {NoStop}%
\bibitem [{\citenamefont {Nimmrichter}(2014)}]{Nimmrichter2013b}%
  \BibitemOpen
  \bibfield  {author} {\bibinfo {author} {\bibfnamefont {S.}~\bibnamefont
  {Nimmrichter}},\ }\href {https://doi.org/10.1007/978-3-319-07097-1} {\emph
  {\bibinfo {title} {Macroscopic Matter Wave Interferometry}}}\ (\bibinfo
  {publisher} {Springer International Publishing},\ \bibinfo {year}
  {2014})\BibitemShut {NoStop}%
\bibitem [{\citenamefont {Belenchia}\ \emph {et~al.}(2019)\citenamefont
  {Belenchia}, \citenamefont {Gasbarri}, \citenamefont {Kaltenbaek},
  \citenamefont {Ulbricht},\ and\ \citenamefont
  {Paternostro}}]{Belenchia2019a}%
  \BibitemOpen
  \bibfield  {author} {\bibinfo {author} {\bibfnamefont {A.}~\bibnamefont
  {Belenchia}}, \bibinfo {author} {\bibfnamefont {G.}~\bibnamefont {Gasbarri}},
  \bibinfo {author} {\bibfnamefont {R.}~\bibnamefont {Kaltenbaek}}, \bibinfo
  {author} {\bibfnamefont {H.}~\bibnamefont {Ulbricht}},\ and\ \bibinfo
  {author} {\bibfnamefont {M.}~\bibnamefont {Paternostro}},\ }\href
  {https://doi.org/10.1103/PhysRevA.100.033813} {\bibfield  {journal} {\bibinfo
   {journal} {Phys. Rev. A}\ }\textbf {\bibinfo {volume} {100}},\ \bibinfo
  {pages} {033813} (\bibinfo {year} {2019})}\BibitemShut {NoStop}%
\bibitem [{\citenamefont {Kaltenbaek}\ \emph {et~al.}(2012)\citenamefont
  {Kaltenbaek}, \citenamefont {Hechenblaikner}, \citenamefont {Kiesel},
  \citenamefont {Romero-Isart}, \citenamefont {Schwab}, \citenamefont
  {Johann},\ and\ \citenamefont {Aspelmeyer}}]{Kaltenbaek2012b}%
  \BibitemOpen
  \bibfield  {author} {\bibinfo {author} {\bibfnamefont {R.}~\bibnamefont
  {Kaltenbaek}}, \bibinfo {author} {\bibfnamefont {G.}~\bibnamefont
  {Hechenblaikner}}, \bibinfo {author} {\bibfnamefont {N.}~\bibnamefont
  {Kiesel}}, \bibinfo {author} {\bibfnamefont {O.}~\bibnamefont
  {Romero-Isart}}, \bibinfo {author} {\bibfnamefont {K.~C.}\ \bibnamefont
  {Schwab}}, \bibinfo {author} {\bibfnamefont {U.}~\bibnamefont {Johann}},\
  and\ \bibinfo {author} {\bibfnamefont {M.}~\bibnamefont {Aspelmeyer}},\
  }\href {https://doi.org/10.1007/s10686-012-9292-3} {\bibfield  {journal}
  {\bibinfo  {journal} {Experimental Astronomy}\ }\textbf {\bibinfo {volume}
  {34}},\ \bibinfo {pages} {123–164} (\bibinfo {year} {2012})}\BibitemShut
  {NoStop}%
\bibitem [{\citenamefont {Stern}\ \emph {et~al.}(1990)\citenamefont {Stern},
  \citenamefont {Aharonov},\ and\ \citenamefont {Imry}}]{Stern1990}%
  \BibitemOpen
  \bibfield  {author} {\bibinfo {author} {\bibfnamefont {A.}~\bibnamefont
  {Stern}}, \bibinfo {author} {\bibfnamefont {Y.}~\bibnamefont {Aharonov}},\
  and\ \bibinfo {author} {\bibfnamefont {Y.}~\bibnamefont {Imry}},\ }\href
  {https://doi.org/10.1103/PhysRevA.41.3436} {\bibfield  {journal} {\bibinfo
  {journal} {Phys. Rev. A}\ }\textbf {\bibinfo {volume} {41}},\ \bibinfo
  {pages} {3436} (\bibinfo {year} {1990})}\BibitemShut {NoStop}%
\bibitem [{\citenamefont {Dalibard}\ \emph {et~al.}(1992)\citenamefont
  {Dalibard}, \citenamefont {Castin},\ and\ \citenamefont
  {M\o{}lmer}}]{Dalibard1992}%
  \BibitemOpen
  \bibfield  {author} {\bibinfo {author} {\bibfnamefont {J.}~\bibnamefont
  {Dalibard}}, \bibinfo {author} {\bibfnamefont {Y.}~\bibnamefont {Castin}},\
  and\ \bibinfo {author} {\bibfnamefont {K.}~\bibnamefont {M\o{}lmer}},\ }\href
  {https://doi.org/10.1103/PhysRevLett.68.580} {\bibfield  {journal} {\bibinfo
  {journal} {Phys. Rev. Lett.}\ }\textbf {\bibinfo {volume} {68}},\ \bibinfo
  {pages} {580} (\bibinfo {year} {1992})}\BibitemShut {NoStop}%
\bibitem [{\citenamefont {Cruickshank}\ \emph {et~al.}(2026)\citenamefont
  {Cruickshank}, \citenamefont {La~Rooij}, \citenamefont {Kerr}, \citenamefont
  {Hilker}, \citenamefont {Kuhr},\ and\ \citenamefont
  {Haller}}]{Cruickshank2026}%
  \BibitemOpen
  \bibfield  {author} {\bibinfo {author} {\bibfnamefont {R.}~\bibnamefont
  {Cruickshank}}, \bibinfo {author} {\bibfnamefont {A.}~\bibnamefont
  {La~Rooij}}, \bibinfo {author} {\bibfnamefont {E.~F.}\ \bibnamefont {Kerr}},
  \bibinfo {author} {\bibfnamefont {T.}~\bibnamefont {Hilker}}, \bibinfo
  {author} {\bibfnamefont {S.}~\bibnamefont {Kuhr}},\ and\ \bibinfo {author}
  {\bibfnamefont {E.}~\bibnamefont {Haller}},\ }\href
  {https://doi.org/10.1364/oe.581265} {\bibfield  {journal} {\bibinfo
  {journal} {Optics Express}\ }\textbf {\bibinfo {volume} {34}},\ \bibinfo
  {pages} {623} (\bibinfo {year} {2026})}\BibitemShut {NoStop}%
\bibitem [{\citenamefont {Case}\ \emph {et~al.}(2009)\citenamefont {Case},
  \citenamefont {Tomandl}, \citenamefont {Deachapunya},\ and\ \citenamefont
  {Arndt}}]{Case2009}%
  \BibitemOpen
  \bibfield  {author} {\bibinfo {author} {\bibfnamefont {W.~B.}\ \bibnamefont
  {Case}}, \bibinfo {author} {\bibfnamefont {M.}~\bibnamefont {Tomandl}},
  \bibinfo {author} {\bibfnamefont {S.}~\bibnamefont {Deachapunya}},\ and\
  \bibinfo {author} {\bibfnamefont {M.}~\bibnamefont {Arndt}},\ }\href
  {https://doi.org/10.1364/oe.17.020966} {\bibfield  {journal} {\bibinfo
  {journal} {Optics Express}\ }\textbf {\bibinfo {volume} {17}},\ \bibinfo
  {pages} {20966} (\bibinfo {year} {2009})}\BibitemShut {NoStop}%
\bibitem [{\citenamefont {Brukner}\ and\ \citenamefont
  {Zeilinger}(1997)}]{Brukner1997}%
  \BibitemOpen
  \bibfield  {author} {\bibinfo {author} {\bibfnamefont {{\v C}.}~\bibnamefont
  {Brukner}}\ and\ \bibinfo {author} {\bibfnamefont {A.}~\bibnamefont
  {Zeilinger}},\ }\href {https://doi.org/10.1103/PhysRevLett.79.2599}
  {\bibfield  {journal} {\bibinfo  {journal} {Phys. Rev. Lett.}\ }\textbf
  {\bibinfo {volume} {79}},\ \bibinfo {pages} {2599} (\bibinfo {year}
  {1997})}\BibitemShut {NoStop}%
\bibitem [{\citenamefont {Santos}\ \emph {et~al.}(2018)\citenamefont {Santos},
  \citenamefont {Castro},\ and\ \citenamefont {Torres}}]{Santos2018}%
  \BibitemOpen
  \bibfield  {author} {\bibinfo {author} {\bibfnamefont {E.~A.}\ \bibnamefont
  {Santos}}, \bibinfo {author} {\bibfnamefont {F.}~\bibnamefont {Castro}},\
  and\ \bibinfo {author} {\bibfnamefont {R.}~\bibnamefont {Torres}},\ }\href
  {https://doi.org/10.1103/PhysRevA.97.043853} {\bibfield  {journal} {\bibinfo
  {journal} {Phys. Rev. A}\ }\textbf {\bibinfo {volume} {97}},\ \bibinfo
  {pages} {043853} (\bibinfo {year} {2018})}\BibitemShut {NoStop}%
\bibitem [{\citenamefont {Deng}\ \emph {et~al.}(1999)\citenamefont {Deng},
  \citenamefont {Hagley}, \citenamefont {Denschlag}, \citenamefont {Simsarian},
  \citenamefont {Edwards}, \citenamefont {Clark}, \citenamefont {Helmerson},
  \citenamefont {Rolston},\ and\ \citenamefont {Phillips}}]{Deng1999}%
  \BibitemOpen
  \bibfield  {author} {\bibinfo {author} {\bibfnamefont {L.}~\bibnamefont
  {Deng}}, \bibinfo {author} {\bibfnamefont {E.~W.}\ \bibnamefont {Hagley}},
  \bibinfo {author} {\bibfnamefont {J.}~\bibnamefont {Denschlag}}, \bibinfo
  {author} {\bibfnamefont {J.~E.}\ \bibnamefont {Simsarian}}, \bibinfo {author}
  {\bibfnamefont {M.}~\bibnamefont {Edwards}}, \bibinfo {author} {\bibfnamefont
  {C.~W.}\ \bibnamefont {Clark}}, \bibinfo {author} {\bibfnamefont
  {K.}~\bibnamefont {Helmerson}}, \bibinfo {author} {\bibfnamefont {S.~L.}\
  \bibnamefont {Rolston}},\ and\ \bibinfo {author} {\bibfnamefont {W.~D.}\
  \bibnamefont {Phillips}},\ }\href
  {https://doi.org/10.1103/PhysRevLett.83.5407} {\bibfield  {journal} {\bibinfo
   {journal} {Phys. Rev. Lett.}\ }\textbf {\bibinfo {volume} {83}},\ \bibinfo
  {pages} {5407} (\bibinfo {year} {1999})}\BibitemShut {NoStop}%
\bibitem [{\citenamefont {Feynman}(1948)}]{Feynman1948}%
  \BibitemOpen
  \bibfield  {author} {\bibinfo {author} {\bibfnamefont {R.~P.}\ \bibnamefont
  {Feynman}},\ }\href {https://doi.org/10.1103/revmodphys.20.367} {\bibfield
  {journal} {\bibinfo  {journal} {Reviews of Modern Physics}\ }\textbf
  {\bibinfo {volume} {20}},\ \bibinfo {pages} {367–387} (\bibinfo {year}
  {1948})}\BibitemShut {NoStop}%
\bibitem [{\citenamefont {Torbatiyan}\ \emph {et~al.}(2026)\citenamefont
  {Torbatiyan}, \citenamefont {Halcı}, \citenamefont {Udovičić},
  \citenamefont {Ulčakar},\ and\ \citenamefont
  {Kaltenbaek}}]{Torbatiyan2026b}%
  \BibitemOpen
  \bibfield  {author} {\bibinfo {author} {\bibfnamefont {H.}~\bibnamefont
  {Torbatiyan}}, \bibinfo {author} {\bibfnamefont {B.}~\bibnamefont {Halcı}},
  \bibinfo {author} {\bibfnamefont {A.}~\bibnamefont {Udovičić}}, \bibinfo
  {author} {\bibfnamefont {L.}~\bibnamefont {Ulčakar}},\ and\ \bibinfo
  {author} {\bibfnamefont {R.}~\bibnamefont {Kaltenbaek}},\ }\href
  {https://doi.org/10.5281/zenodo.23057278} {10.5281/zenodo.23057278} (\bibinfo
  {year} {2026})\BibitemShut {NoStop}%
\bibitem [{\citenamefont {Carmichael}(1993)}]{Carmichael1993}%
  \BibitemOpen
  \bibfield  {author} {\bibinfo {author} {\bibfnamefont {H.}~\bibnamefont
  {Carmichael}},\ }\href {https://doi.org/10.1007/978-3-540-47620-7_8} {\emph
  {\bibinfo {title} {An Open Systems Approach to Quantum Optics}}}\ (\bibinfo
  {publisher} {Springer Berlin Heidelberg},\ \bibinfo {address} {Berlin,
  Heidelberg},\ \bibinfo {year} {1993})\ pp.\ \bibinfo {pages}
  {113--125}\BibitemShut {NoStop}%
\bibitem [{\citenamefont {Buralli}\ and\ \citenamefont
  {Morris}(1992)}]{Buralli1992}%
  \BibitemOpen
  \bibfield  {author} {\bibinfo {author} {\bibfnamefont {D.~A.}\ \bibnamefont
  {Buralli}}\ and\ \bibinfo {author} {\bibfnamefont {G.~M.}\ \bibnamefont
  {Morris}},\ }\href {https://doi.org/10.1364/ao.31.004389} {\bibfield
  {journal} {\bibinfo  {journal} {Applied Optics}\ }\textbf {\bibinfo {volume}
  {31}},\ \bibinfo {pages} {4389} (\bibinfo {year} {1992})}\BibitemShut
  {NoStop}%
\bibitem [{\citenamefont {Moreno}\ \emph {et~al.}(2011)\citenamefont {Moreno},
  \citenamefont {Marquez}, \citenamefont {Iemmi}, \citenamefont {Campos},\ and\
  \citenamefont {Yzuel}}]{Moreno2011}%
  \BibitemOpen
  \bibfield  {author} {\bibinfo {author} {\bibfnamefont {I.}~\bibnamefont
  {Moreno}}, \bibinfo {author} {\bibfnamefont {A.}~\bibnamefont {Marquez}},
  \bibinfo {author} {\bibfnamefont {C.}~\bibnamefont {Iemmi}}, \bibinfo
  {author} {\bibfnamefont {J.}~\bibnamefont {Campos}},\ and\ \bibinfo {author}
  {\bibfnamefont {M.~J.}\ \bibnamefont {Yzuel}},\ }in\ \href
  {https://doi.org/10.1109/wio.2011.5981461} {\emph {\bibinfo {booktitle} {2011
  10th Euro-American Workshop on Information Optics}}}\ (\bibinfo  {publisher}
  {IEEE},\ \bibinfo {year} {2011})\ p.\ \bibinfo {pages} {1–4}\BibitemShut
  {NoStop}%
\end{thebibliography}%

\end{document}